\documentclass[10pt,journal,compsoc]{IEEEtran}
\usepackage{color}
\usepackage{xcolor}

\ifCLASSOPTIONcompsoc
\usepackage[nocompress]{cite}
\else
\usepackage{cite}
\fi

\usepackage[utf8]{inputenc}
\usepackage{longtable}
\usepackage{graphicx}
\usepackage{color}

\usepackage[cmex10]{amsmath}

\usepackage{array}
\usepackage{mdwmath}
\usepackage{mdwtab}
\usepackage{subfig}
\usepackage{url}
\usepackage{bm}
\usepackage{hyperref}
\usepackage{wrapfig,booktabs}
\usepackage{multirow}
\usepackage{booktabs}
\usepackage{amssymb}
\usepackage{lscape}

\usepackage{rotating}
\usepackage{cleveref}

\usepackage{tikz}
\usepackage{pifont}

\usepackage{nicefrac}

\usepackage{algorithm}
\usepackage{algorithmic}

\graphicspath{{Figures/}}
\DeclareGraphicsExtensions{.pdf,.eps}
\newcommand\norm[1]{\left\lVert#1\right\rVert}

\newcommand{\branch}{\beta}
\newcommand{\curve}{f}
\newcommand{\threeDTree}{T}
\newcommand{\fourDTree}{H}

\newcommand{\srvf}{q}
\newcommand{\pcasrvf}{w}
\newcommand{\srvfbranch}{\srvf}
\newcommand{\srvfthreeDTree}{Q}
\newcommand{\srvffourDTree}{\pcasrvf} 

\newcommand{\preshapespace}{\mathcal{C}}
\newcommand{\preshapespaceThreeDTreees}{\preshapespace_\threeDTree}
\newcommand{\preshapesrvts}{\preshapespace_\srvfthreeDTree}

\newcommand{\pcaspace}{\preshapespace_{\text{PCA}}}
\newcommand{\srvfpcaspace}{\preshapespace_\pcasrvf}

\newcommand{\rotation}{O}
\newcommand{\rotationspace}{SO(3)}
\newcommand{\reparm}{\gamma}
\newcommand{\diffeotree}{\boldsymbol{\reparm}}
\newcommand{\reparmspace}{\Gamma}
\newcommand{\permutation}{\sigma}

\newcommand{\permutetree}{\boldsymbol{\permutation}}

\newcommand{\fourDcurve}{\alpha}
\newcommand{\preshapespaceFourDcurve}{\preshapespace_\fourDcurve}
\newcommand{\fourDcurveinpca}{\alpha^{\text{pca}}}
\newcommand{\diffalpha}{\alpha'^{{\text{pca}}}(t)}
\newcommand{\fourDcurveinpcaone}{\fourDcurveinpca_1}
\newcommand{\fourDcurveinpcatwo}{\fourDcurveinpca_2}

\newcommand{\covMatrix}{K}

\newcommand{\reparmcurve}{\xi}   
\newcommand{\reparmcurvespace}{\Xi} 

\newcommand{\optimal}{^*}

\newcommand{\totalthreeDtree}{m}
\newcommand{\totalfourDtree}{n}

\newcommand{\meantree}{\boldsymbol{\mu}}
\newcommand{\meansrvft}{\boldsymbol{\mu}_{\srvfbranch}}

\def\argmin{\mathop{\rm argmin}}

\newcommand{\ie}{{i.e., }}
\newcommand{\eg}{{e.g., }}
\newcommand{\etal}{{et al.}}

\newcommand{\ltwo}{\mathbb{L}^2}
\newcommand{\real}{\mathbb{R}}
\newcommand{\rthree}{\real^3}

\newcommand{\rpositive}{\real^+}
\newcommand{\noi}{\noindent}

\newcommand{\bifurcation}{s}

\begin{document}
\bstctlcite{IEEEexample:BSTcontrol}




\title{A Riemannian Framework for the Elastic  Analysis  of the Spatiotemporal Variability in the Shape and Structure of Tree-like 4D Objects}

\author{Tahmina Khanam, Hamid Laga, Mohammed Bennamoun, Guanjin Wang, Ferdous Sohel, Farid Boussaid, Guan Wang, Anuj Srivastava
	
	\IEEEcompsocitemizethanks{\IEEEcompsocthanksitem Tahmina Khanam is with Murdoch University. Email: 34719017@student.murdoch.edu.au%
		\IEEEcompsocthanksitem  Hamid Laga, Guanjin Wang, and Ferdous Sohel are with Murdoch University. Emails: \{H.Laga,guanjin.wang,f.sohel\}@murdoch.edu.au%
        \IEEEcompsocthanksitem Mohammed Bennamoun and Farid Boussaid are with the University of Western Australia. Emails: \{mohammed.bennamoun,farid.boussaid\}@uwa.edu.au%
        \IEEEcompsocthanksitem Guan Wang is with Huzhou Normal University. Email: wangguan12621@gmail.com
		\IEEEcompsocthanksitem Anuj Srivastava is with Johns Hopkins University. Email: anuj.srivastava@jhu.edu%
	}
	\thanks{Manuscript received XXX; revised yyy.}
}

\markboth{}%
{Name \MakeLowercase{\etalnospace}: title}

\IEEEcompsoctitleabstractindextext{
\begin{abstract} 
This paper introduces a novel computational framework for modeling and analyzing the spatiotemporal shape variability of tree-like 4D objects—three-dimensional structures whose shapes deform and evolve over time. Tree-like 3D objects, such as botanical trees and plants, deform and grow at different rates. In this process, they bend and stretch their branches and change their branching structure, making their spatiotemporal registration challenging. We address this problem within a Riemannian framework that represents tree-like 3D objects as points in a tree-shape space endowed with a proper elastic metric that quantifies branch bending, stretching, and topological changes.
With this setting, a 4D tree-like object becomes a trajectory in the tree-shape space. Thus, the problem of modeling and analyzing the spatiotemporal variability in tree-like 4D objects reduces to the analysis of trajectories within this tree-shape space.
However, performing spatiotemporal registration and subsequently computing geodesics and statistics in the nonlinear tree-shape space is inherently challenging, as these tasks rely on complex nonlinear optimizations.  Our core contribution is the mapping of the tree-like 3D objects to the space of the Extended Square Root Velocity Field (ESRVF)~\cite{srivastava2010shape} where the complex elastic metric is reduced to the $\ltwo$ metric. By solving spatial registration in the ESRVF space,  analyzing tree-like 4D objects can be reformulated as the problem of analyzing elastic trajectories in the ESRVF space.
Based on this formulation, we develop a comprehensive framework for analyzing the spatiotemporal dynamics of tree-like objects, including registration under large deformations and topological differences, geodesic computation, statistical summarization through mean trajectories and modes of variation, and the synthesis of new, random tree-like 4D shapes.
We demonstrate the effectiveness of the proposed framework on both synthetic 4D botanical tree models and real plant data, illustrating its robustness and practical applicability in capturing complex spatiotemporal shape evolution. Source code is publicly available in our \href{https://github.com/Tahmina979/4Dtree_shape_with_geometry}{GitHub} repository. 
\end{abstract}

}

\maketitle

\IEEEdisplaynotcompsoctitleabstractindextext

\IEEEpeerreviewmaketitle

\section{Introduction}
\label{sec:introduction}

\IEEEPARstart{S}{hape} is a  fundamental property of  every 3D object, whether it is natural or man-made. However, real-world objects are rarely static—they move, interact with their surroundings, and evolve over time, often exhibiting complex growth behaviors. These dynamic processes frequently result in complex non-rigid deformations and structural changes.  The ability to \textbf{(1)} mathematically characterize and model typical patterns of shape deformation and their variability, both within and across object classes, and \textbf{(2)} develop generative models for such spatiotemporal deformations, is central to applications across biology, computer vision, and computer graphics. For example, modeling the growth and deformation of biological structures such as plants or trees can yield insights into how environmental factors influence development. Similarly, analyzing dynamic changes in anatomical structures such as blood vessels or airway trees may help distinguish between changes due to normal aging and those associated with pathological conditions. 

This paper presents a novel framework for the statistical analysis of the spatiotemporal deformation patterns in dynamic 3D shapes, hereinafter referred to as 4D (or 3D $+$ time) shapes.  Although increasing attention has been given to 4D shape analysis in recent years, most prior work has focused on reconstructing~\cite{jiang2022lord,jiang2022h4d} such shapes or modeling deformations that preserve the original topology and branching structure~\cite{laga20224d}. In contrast, this work addresses a broader class of tree-like 4D shapes that not only undergo elastic bending and stretching but also experience changes in their branching topology over time.
Given a set of such 4D objects, our goal is to:
\begin{enumerate}

    \item Perform spatiotemporal registration of the 4D tree shapes. Spatial registration refers to the process of finding one-to-one branch-wise and surface-based point-wise correspondences. Temporal registration refers to the process of temporally aligning 4D tree-like shapes, which may deform and grow at varying rates.

    \item Compute average deformation patterns across a population of 4D tree-like shapes, \eg to capture typical growth trajectories of plants or trees under varying conditions;

    \item Extract principal modes of variation, analogous to Principal Component Analysis (PCA), extended to model variability in 4D tree-like shapes; and
    
    \item Develop generative models that can synthesize new tree-like 4D shapes either randomly or under specific constraints, enabling applications in simulation and virtual environment generation.
     
\end{enumerate}

\noi Achieving these objectives poses significant challenges. Tree-like 3D shapes often undergo large elastic deformations and exhibit substantial variability in their branching structure, making spatial correspondence estimation highly nontrivial. When extended to 4D, additional complexity arises from temporal variability, particularly due to inconsistent growth rates across samples. Effective analysis thus requires robust and efficient spatiotemporal registration methods.

A key contribution of this paper is the representation of tree-like 3D shapes in the Extended Square Root Velocity Field (ESRVF) space~\cite{srivastava2010shape,wang2023elastic}. This representation simplifies the complex elastic metric into a standard $\ltwo$ metric, enabling more tractable spatial registration. Once mapped to this space, 4D tree-like shapes can be modeled as elastic trajectories, reducing the problem of spatiotemporal analysis to that of analyzing these trajectories in ESRVF space. Building on previous work~\cite{wang2023elastic}, which developed tools for spatial registration of tree-like 3D shapes using ESRVF, we extend the approach to 4D and introduce a comprehensive framework for: spatiotemporal registration, geodesic computation, statistical summarization, and generative modeling of tree-like 4D shapes.

\begin{figure}[t]
    \centering
    \includegraphics[width=\linewidth]{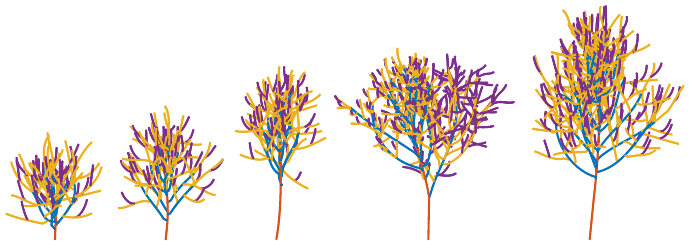}\\
    \small{\textbf{(a)}}
    \\
    \includegraphics[width=\linewidth]{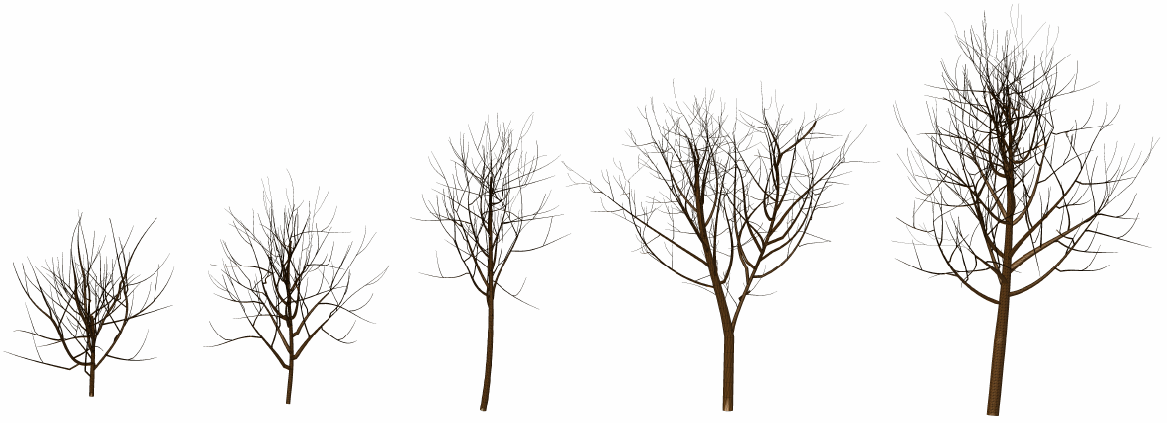}
    \small{\textbf{(b)}}
    
    \caption{Visual comparison between \textbf{(a)} the skeletal representation of 4D tree-like shapes used in our earlier conference paper~\cite{khanam2024riemannian}, and \textbf{(b)} the full 3D geometry considered in the current work. In \textbf{(a)}, trees are represented by layered skeletal structures without explicit branch geometry. Colors indicate branch hierarchy: 
    \textcolor{red}{red}  indicates the main branch, while the \textcolor{blue}{blue},  \textcolor{yellow}{yellow}, and \textcolor{purple}{purple} correspond, respectively,   to the \textcolor{blue}{$2^{nd}$},  \textcolor{yellow}{$3^{rd}$}, and \textcolor{purple}{$4^{th}$} layers of branches.
    In contrast, \textbf{(b)} incorporates complete geometric detail of the branches, enabling a richer and more realistic modeling of 4D shape evolution.} 
    \label{fig:tree_vis}
\end{figure}

Prior ESRVF-based works~\cite{wang2023elastic} addressed only static 3D tree-like objects and did not model temporal variability, temporal registration, or 4D statistical analysis. This work builds upon our earlier conference paper~\cite{khanam2024riemannian}, which introduced a preliminary version of the framework focused solely on skeletal representations of tree-like 4D shapes, entirely ignoring the 3D geometry of the branches. In contrast, the current paper extends both the ESRVF representation and associated computational tools to capture complete 4D geometry and structural details. Fig.~\ref{fig:tree_vis} illustrates the key representational differences between this extended work and the earlier version.

The rest of the paper is organized as follows. Section~\ref{sec:related_work} reviews the related work. Section~\ref{sec:3D_shape_space} presents the proposed ESRVF representation of tree-like 3D shapes. Section~\ref{sec:4D_tree_shape_space} generalizes the representation to 4D tree-like shapes and details the proposed computational tools for the spatiotemporal registration and geodesics computation between 4D tree-shapes. Section~\ref{sec:statistical_analysis} describes methods for statistical summarization and synthesis of new shapes. Section~\ref{sec:result} presents experimental results. Section~\ref{sec:conclusion} concludes the paper.

\section{Related works}
\label{sec:related_work}
This section reviews relevant prior work on the analysis of elastic 3D shapes (Section~\ref{sec:related_work_3D_shapes}) and dynamic 4D shapes (Section~\ref{sec:related_work_4D_shapes}).
\subsection{3D shape analysis}
\label{sec:related_work_3D_shapes}
Early research in shape analysis focused on 3D objects that undergo bending and stretching while preserving their underlying topology and structure. Spatial correspondence between such shapes was traditionally achieved using methods like non-rigid Iterated Closest Point (ICP)~\cite{besl1992method,zhang2021fast,tsumura2023body}, often combined with hand-crafted feature matching techniques~\cite{guo2013trisi,masuda2009log,stamos2003automated,yamany2002surface,bariya20123d,malassiotis2007snapshots,tombari2010unique}. Once correspondences are established and shapes are aligned, the shapes can be treated as points in a high-dimensional shape space, where paths between them represent deformation trajectories. Equipping this space with an appropriate metric allows for quantifying the amount of deformation required to align one shape with another. In this context, geodesics—shortest paths in the shape space—correspond to optimal deformation paths under the given metric. A comprehensive overview of the various shape spaces and associated metrics can be found in Laga \etal~\cite{laga2018survey}.
Initial works~\cite{kendall1977diffusion,le1993riemannian,dryden2016statistical} use the $\ltwo$ metric, which is only suitable for rigid shapes and shapes that undergo small nonrigid deformations.  Later, more expressive metrics have been introduced to capture and quantify large nonrigid deformations. For example, Kilian \etal~\cite{kilian2007geometric}  compute geodesics by penalizing deviations from rigidity and isometry.  Drawing from the elasticity theory in physics, Wirth \etal~\cite{wirth2011continuum} and  Zhang \etal~\cite{zhang2015shell} modeled stretching using the Cauchy-Green strain tensor and bending via differences in shape operators. Windheuser \etal~\cite{windheuser2011geometrically}  measured bending through variations in the mean curvature, while  Heeren \etal~\cite{heeren2012time}  used differences in the second fundamental form for bending and changes in the first fundamental form for stretching.

Although these physically motivated metrics are highly expressive, they tend to be computationally intensive. This is because correspondence estimation, geodesic computation, and statistical summarization all require solving complex nonlinear optimization problems. A common workaround is to linearize the shape space by mapping shapes to the tangent space at a reference shape, typically the mean, which is Euclidean. This allows the use of standard PCA to compute shape means and modes of variation~\cite{fletcher2004principal}. However, such tangent-space approximations are only valid for shapes undergoing small deformations relative to the base shape.

Jermyn \etal~\cite{jermyn2012elastic} introduced a more efficient approach by showing that the $\ltwo$ metric in the Square Root Normal Field (SRNF) space approximates an elastic metric. Specifically, it provides a weighted combination of bending and stretching, reducing computational overhead while maintaining a reasonable approximation of elastic deformation. Nonetheless, this representation is not designed to capture topological or structural changes, which are central to the analysis of tree-like 3D shapes.

Unlike generic surfaces, tree-like 3D shapes such as botanical trees and plants not only bend and stretch but also exhibit dynamic changes in their branching structure. To address such structural variability, Billera \etal~\cite{billera2001geometry} and Owen \etal~\cite{owen2010fast} introduced the concept of continuous tree spaces, which model changes in topology but ignore geometric branch information. Feragen \etal~\cite{feragen2010geometries,feragen2011means,feragen2012toward,feragen2013tree} later proposed tree-shape spaces and corresponding metrics to support statistical analysis of tree-like shapes. However, their methods are constrained to relatively simple branching structures due to the high computational cost of establishing correspondences.
To mitigate this, Wang \etal~\cite{wang2018shape,wang2018statistical} proposed pre-computing correspondences before performing statistical analysis. Their framework, based on the Quotient Euclidean Distance (QED), handles topological variation through operations like edge collapse and node split. However, when applied to shapes with significant structural differences, this approach can result in excessive shrinkage along the geodesic path. Duncan \etal~\cite{duncan2018statistical} proposed an alternative that measures topological changes in terms of branch sliding along parent branches, but their method is limited to skeletal representations.
Guan \etal~\cite{wang2023elastic} extended these ideas by incorporating full 3D branch geometry into the representation, enabling more comprehensive modeling of elastic deformations and structural variability. Building upon this, the current work generalizes the ESRVF-based representation to the 4D setting, allowing for the modeling and analysis of spatiotemporal variability in tree-like 3D shapes over time.

\subsection{4D shape analysis}
\label{sec:related_work_4D_shapes}

Most existing work on 4D shape analysis~\cite{lee2024tree,wand2007reconstruction,beeler2011high,tevs2012animation,li2017learning} has primarily focused on the reconstruction of 3D objects that deform over time. While such efforts have advanced dynamic reconstruction, the broader objective of 4D shape analysis is to model both spatial and temporal variability across multiple deforming shapes—particularly when those shapes grow or evolve at different rates.
Early efforts in this area~\cite{anirudh2015elastic,amor2015action} concentrated on longitudinal analysis of 2D shapes. Only recently have researchers begun to tackle the full complexity of 4D shapes. Laga \etal~\cite{laga20224d}, for instance, introduced a Riemannian framework along with computational tools to analyze the spatiotemporal variability of genus-0 4D surfaces with fixed topology. Their approach models a 4D shape as a trajectory in a high-dimensional Riemannian shape space, using the Square Root Normal Field (SRNF) representation~\cite{jermyn2012elastic,jermyn2017elastic} for spatial registration and the Transported SRVF for temporal alignment. However, their method assumes fixed topology and thus cannot accommodate topological changes such as those found in tree-like structures—an issue this paper specifically addresses.
Other methods, such as those by Debavelaere \etal~\cite{debavelaere2020learning} and Bone \etal~\cite{bone2020learning}, use Large Deformation Diffeomorphic Metric Mapping (LDDMM) to represent a 4D shape as a time-evolving deformation of a 3D volume. They compute geodesics in this Riemannian setting by modeling shape evolution as a volumetric flow. Although this approach can capture complex deformations, it is computationally and memory-intensive due to its reliance on volumetric representations.
To the best of our knowledge, existing work that specifically addresses tree-like 4D shapes has focused solely on spatial registration~\cite{zhang2023spatio,wang2022plantmove,lobefaro2023estimating,magistri2020segmentation,chebrolu2020spatio}. For example,~\cite{magistri2020segmentation,chebrolu2020spatio} register time series of growing 3D plants by extracting and matching key points across frames. However, these methods do not perform temporal registration, nor do they offer metrics or algorithms for computing geodesics or deriving statistical summaries in the 4D domain.

This paper builds upon the representation and computational tools proposed in our earlier conference work~\cite{khanam2024riemannian}, which was limited to the skeletal structure of tree-like 4D shapes. In contrast, the current framework generalizes that approach to incorporate full 3D geometry, allowing for a richer and more accurate analysis of spatiotemporal variability in tree-like 4D objects.

\section{Shape space of 3D tree-like structures}
\label{sec:3D_shape_space}

\begin{figure}[!t]
    \centering
    \includegraphics[width=\linewidth]{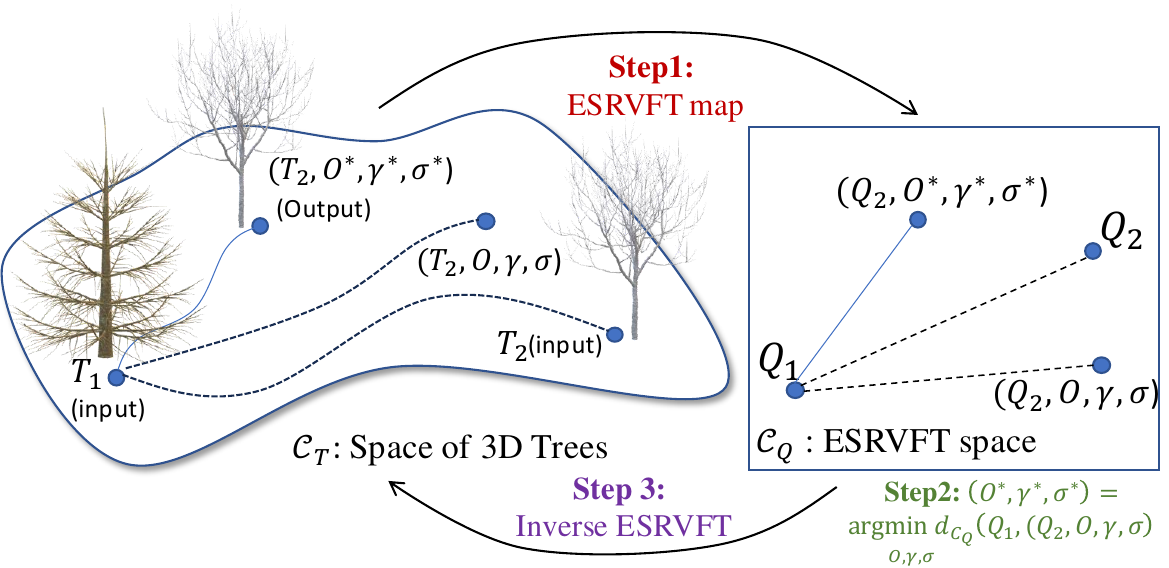}
    \caption{Overview of the proposed spatial registration framework for tree-like 3D shapes. The method operates in three steps: Step 1: Both input shapes $\threeDTree_1$ and $\threeDTree_2$ are mapped to the ESRVFT space $\preshapesrvts$ via the ESRVFT map. Step 2: Optimal alignment parameters $(\rotation\optimal,  \diffeotree\optimal, \permutetree\optimal)$ are computed by minimizing the distance between the transformed shapes in $\preshapesrvts$, \ie $\argmin_{\rotation, \diffeotree,\permutetree} d_{\preshapesrvts}\left(\srvfthreeDTree_1, (\srvfthreeDTree_2, \rotation, \diffeotree, \permutetree) \right)$. Step 3: The aligned shape is then mapped back to the original tree space $\preshapespaceThreeDTreees$ using the inverse ESRVFT transform, producing the spatially registered output $(\threeDTree_2, \rotation\optimal,  \diffeotree\optimal, \permutetree\optimal)$.}
    \label{fig:spatial registration}
\end{figure}

We first describe the mathematical framework we propose to represent the 3D geometry and structure of tree-like 3D shapes (Section~\ref{sec:representation}). We will then introduce an elastic Riemannian metric for quantifying bending, stretching, and structural changes in tree-like 3D shapes (Section~\ref{sect:srvf_3Dtree}). Finally, we propose in Section~\ref{sect:spatial_registraion_within} a set of computational tools for the joint spatial registration and geodesics computation between tree-like 3D shapes. 

\subsection{Representation}
\label{sec:representation}
We represent a branch $ \branch$ of a 3D tree shape $\threeDTree$ using its arc-length parameterized skeletal curve, augmented with the thickness of the branch at each point along the curve:
\begin{equation}
\label{eq:beta_rep}
\begin{split} 
 &\branch: [0, 1] \to \rthree \times \rpositive, \\
  &\branch(s) = (\curve(s),r(s)) = (x(s), y(s), z(s),r(s)).
\end{split}   
\end{equation}

\noi  Here, $(x(s), y(s), z(s)) \in \rthree$ defines a point along the skeletal curve of the branch and $r(s) \in \rpositive$ is the thickness of the branch at that point. We structure a 3D tree into layers:
\begin{itemize}
    \item The first layer $\branch^0$ is composed of the main branch and  $k^0$ sub-trees $\{\threeDTree^1_i\}_{i=1}^{k^0}$ attached to it at  bifurcation points $\{\bifurcation^0_i\}_{i=1}^{k^0}$. Thus, $\threeDTree = (\branch^0, \{\threeDTree^1_{i}, \bifurcation^0_{i}\}_{i=1}^{k^0}) 
$. 
    \item We recursively represent a 3D subtree $\threeDTree^l$ at level $l$ of the hierarchy  as $ \threeDTree^l = (\branch^l,\{\threeDTree^{l+1}_i,\bifurcation^{l}_i\}_{i=1}^{k^l}), \text{ for } l = 1, \dots, L-1$. Here, $L$ is the total number of layers,  $\branch^l$ is the main branch of  $\threeDTree^l$, $k^l$ is the number of subtrees attached to the branch $\branch^l$, and $s^{l}_i \in [0, 1]$ is a bifurcation point on $\branch^l$ of  $\threeDTree^l$. 
\end{itemize}

\begin{figure}[t]
    \centering
    \includegraphics[width=0.2\textwidth]{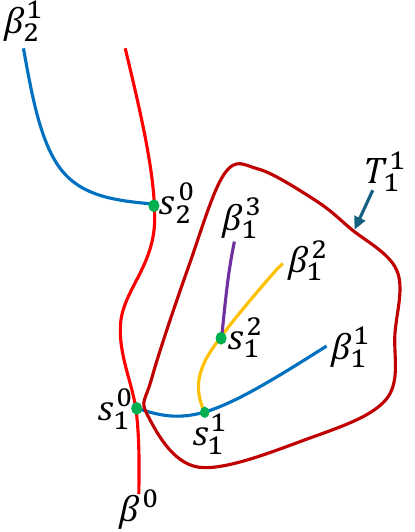}
    \caption{The hierarchical representation of a 3D tree model as defined in Section~\ref{sec:representation}. Here, $\threeDTree$ denotes a tree/subtree, $\branch$ denotes branches and $\bifurcation$ denotes the bifurcation points. The indexing reflects the hierarchical organization: $\branch^0$ is the main trunk, and higher-order branches (\eg $\branch_1^2$,$\branch_1^3$) emerge from successive bifurcations at points $\bifurcation_1^0$, $\bifurcation_1^1$, $\bifurcation_1^2$, etc. The subtree $\threeDTree_1^1$ includes all branches descending from $\bifurcation_1^0$.}
    \label{fig:tree_representation}
\end{figure}

\noi The branches at different levels can have any arbitrary number of subtrees attached to them.  Fig.~\ref{fig:tree_representation} illustrates this layered representation.

A good representation should be invariant to similarity-preserving transformations, \ie translation, scale, rotation, and reparameterization. To achieve translation invariance, we translate each tree so that the start point of its main branch is located at the origin. Scale invariance is optional. For instance, if the goal is to model growth,  then the (relative) scale of the trees needs to be preserved. In our implementation, we perform scale normalization independently within each 4D tree. Specifically, all 3D tree instances within a sequence are normalized relative to the smallest tree in that sequence. We compute the radius of the smallest tree in a 4D tree and then scale all the 3D trees in that 4D tree by this radius. This preserves the relative growth patterns over time while removing unnecessary global scale variations across samples. The invariance to rotation and reparameterization is dealt with at the metric level; see \Cref{sect:srvf_3Dtree} for details. 

We refer to the space of all trees that are normalized for translation, and optionally for scale, as the \emph{pre-tree-shape space}, which we denote by $\preshapespaceThreeDTreees$.  With this representation, a 3D tree becomes a point in the pre-tree shape space $\preshapespaceThreeDTreees$. A 4D tree $\fourDTree$ can then be seen as a time-parameterized curve $\fourDTree: [0, 1] \to \preshapespaceThreeDTreees$ where $[0, 1] $ is the time domain.

\subsection{Spatial registration and geodesics}
\label{sect:srvf_3Dtree}

\subsubsection{The elastic metric for tree-shaped structures}
Consider two 3D tree structures that have the same number of branches. In that case, one can quantify the dissimilarity between the two trees by measuring the amount of bending,  stretching, and branch sliding that one needs to apply to the branches of one tree to align it with the other.  Bending can be measured using changes in the orientation of the tangent vectors to the skeletal curve $\curve$ of the branch $\branch$ along the deformation path. Branch stretching, on the other hand, can be decomposed into two components. The \textbf{first} one is related to the elongation of the branches and can be measured in terms of changes in the magnitude of the tangent vector to the skeletal curve $\curve$ of the branch $\branch$. The \textbf{second} one is related to changes in the thickness of the branches. 

Using these physical quantities would result in a complex elastic metric that is computationally expensive to evaluate. Thus, we propose to represent a tree branch using the Square Root Velocity Function (SRVF)~\cite{srivastava2010shape} of its skeletal curve, augmented with the radius at each point along the skeleton. The main advantage is that the $\ltwo$ metric in the space of SRVFs is equivalent to an elastic metric~\cite{srivastava2010shape}. Thus, one can perform elastic shape analysis using the $\ltwo$ metric in the space of SRVFs. 
Mathematically, the SRVF  $\srvfbranch$ of a curve $\curve$ is defined as $
    \srvfbranch(\curve)(s) = 
         \frac{\curve'(s)}{\sqrt{\|\curve'(s)\|}},     \text{ if } \|\curve'(s)\|\neq 0, \text{ and } 0         \text{ otherwise}$.
To take into account the thickness of the branch $\branch(s)$,  we extend this formulation by following~\cite{wang2023elastic} to include branch thickness and refer to it as the  Extended SRVF (ESRVF). In other words, we define the ESRVF of a tree branch $\branch = (\curve, r) $ as
\begin{equation}
\label{eq:esrvf_Rep}
        \srvfbranch(\branch)(s)=\left(\frac{\curve'(s)}{\sqrt{\|\curve'(s)\|}}, r(s)\right).
\end{equation}

\noi Its main property is that the $\ltwo$ metric in the ESRVF space is equivalent to a weighted sum of bending and stretching in the original space. It is also invertible, up to translation. In other words, given an ESRVF, one can retrieve, up to translation, its corresponding original 3D branch. This significantly simplifies the analysis tasks: instead of measuring the similarity between two 3D branches using the complex elastic metric, one can map them to the ESRVF space, perform the analysis there using the $\ltwo$ metric, and then map the results back to the original space. 

Let $\srvfthreeDTree$ be the ESRVF Tree (ESRVFT) of an entire 3D tree $\threeDTree$ defined by computing the ESRVF of each of its branches and appending their location with respect to their parent branch. We refer to the space of all such trees as the pre-shape space of ESRVFs, denoted by $\preshapesrvts$.

A proper metric on that space $\srvfthreeDTree$ needs to be invariant to the global rotation $\rotation \in \rotationspace$, reparameterization $\diffeotree$ of the branches, and permutations $\permutetree$ of the orders of the lateral subtrees attached to a branch. We define $\diffeotree$ and $\permutetree$ recursively, \ie $\diffeotree = (\reparm^0, \{\diffeotree^{i}\}_{i=1}^{k^0})$ is the reparameterization of the main branch and its subtrees.  Specifically,  $\reparm^0 \in \reparmspace$ is a diffeomorphism that applies to the main branch of $\srvfbranch$ and $\diffeotree^{i}$ is the reparameterization, defined recursively, of the $i-$th subtree attached to the main branch. $\permutetree = (\permutation^0, \{ \permutetree^{i}\}_{i=1}^{k^0})$ where $\permutation^0 \in \permutetree$ is the permutation of the orders of the lateral subtrees on their corresponding main branch $\srvfbranch^0$, and $\permutetree^{i}$ defines recursively these permutations for the $i-$th subtree. Following~\cite{wang2023elastic,duncan2018statistical}, we define a rotation, reparameterization, and index permutation-invariant distance between two 3D trees, represented by their ESRVFTs $\srvfthreeDTree_1$ and $\srvfthreeDTree_2$,  as the infimum over all possible rotations, branch reparameterization, and branch index permutations:
\begin{equation}
	\label{eq:invariantdistance}
 \small{
	\begin{aligned}
		d\left(\srvfthreeDTree_1, \srvfthreeDTree_2\right) = \inf_{\tiny{\begin{tabular}{c} 
							 	$\rotation, \diffeotree , \permutetree $
							\end{tabular}
							}}  d_{\preshapesrvts}\left(\srvfthreeDTree_1, (\srvfthreeDTree_2,  \rotation, \diffeotree, \permutetree\right)), \text{ and }
	\end{aligned}
 }
\end{equation}\vspace{-6pt}
\begin{equation}
	\label{eq:invariant_metric}
\small{
    \begin{split}
	d_{\preshapesrvts}\left(\srvfthreeDTree_1, (\srvfthreeDTree_2,  \rotation, \diffeotree, \permutetree) \right)  =& \lambda_m \parallel\srvfbranch^0_{1} - \rotation(\srvfbranch^0_{2}, \reparm^0) \parallel^2  + 	\\			
			 	\lambda_p \sum_{i=1}^{k^l} \left({s}^{i}_{1} - {s}^{\permutetree(i)}_{2}\right)^2 
+ 
        & \lambda_s \sum_{i=1}^{k^l} d \left( \srvfthreeDTree_1^i, (\srvfthreeDTree_2^{\permutetree(i)},  \rotation, \diffeotree, \permutetree)   \right).
    \end{split}
	}   
\end{equation}

\noi where $k^l$ is the number of side branches at a layer $l=1,\dots, L-1$, and $L$ is the total number of layers. The parameters ($\lambda_m,\lambda_s,\lambda_p$) control the relative cost of deforming the main branch, deforming the subtrees connected to the main branch, and sliding along the main branch the subtrees connected to it, respectively. The parameter $\lambda_m$ controls the amount of bending and stretching along the main branch. $\lambda_p$ controls the amount of branch sliding and $\lambda_s$ controls the contribution of subtree similarity to the overall distance. A high $\lambda_p$ will favor correspondences between spatially nearby side branches, a high $\lambda_m$ will allow greater geometric deformation instead of preserving side branch proximity, and a high $\lambda_s$ will prioritize subtree similarity.

To enable correspondence estimation between trees with different branching structures, we augment the hierarchical representation with null branches whenever a corresponding branch is absent in one of the trees at a given hierarchical level. This produces structurally compatible tree representations, allowing the recursive elastic metric and subtree permutation optimization to operate consistently across trees with different branch counts. During spatial registration, existing branches are therefore matched either to corresponding physical branches or, when no correspondence exists, to null branches introduced for topology balancing. These null branches also provide the mechanism for modelling the appearance of new branches.

\subsubsection{Spatial registration and geodesics} 
\label{sect:spatial_registraion_within}

With this formulation, the spatial registration, \ie the process of establishing branch-wise and point-wise correspondences, between two 3D trees represented using their ESRVFs $\srvfthreeDTree_1$ and $\srvfthreeDTree_2$, can be formulated as that of finding optimal rotation $\rotation\in \rotationspace$, diffeomorphism $\reparm \in \reparmspace$, and permutations $\permutetree$ that align $\srvfthreeDTree_2$ onto $\srvfthreeDTree_1$. This can be formulated as an optimization problem of the form:
\begin{equation}
	\label{eq:registration}
	(\rotation\optimal,  \diffeotree\optimal, \permutetree\optimal) = \argmin_{\rotation, \diffeotree,\permutetree} d_{\preshapesrvts}\left(\srvfthreeDTree_1, (\srvfthreeDTree_2, \rotation,   \diffeotree, \permutetree) \right).
\end{equation}

\noi We solve this optimization problem following the approach described in~\cite{duncan2018statistical,wang2023elastic}. In practice, to spatially register all the instances within a  4D tree,  we proceed sequentially where each 3D tree-shape within a sequence is aligned to the next 3D tree-shape in the sequence, using Eqn.~\ref{eq:registration}. 

With this formulation, computing a geodesic, or the optimal deformation that aligns  $\srvfthreeDTree_2$ onto  $\srvfthreeDTree_1$ is straightforward. Let $\srvfthreeDTree\optimal_2 = (\srvfthreeDTree_2, \rotation\optimal, \diffeotree\optimal, \permutetree\optimal)$ be the aligned tree onto $\srvfthreeDTree_1$. Since the metric is a weighted norm of $\ltwo$ distances, the geodesic $g\optimal$ between $\srvfthreeDTree_1$ and $\srvfthreeDTree\optimal_2$ is the straight line that connects them, \ie $g\optimal(t) = (1 - t)  \srvfthreeDTree_1 + t \srvfthreeDTree\optimal_2, t\in [0, 1]$. Note that ESRVF has two components: the SRVF of the branch skeleton and the thickness; see~\Cref{eq:esrvf_Rep}. Since both are elements of $\ltwo$ (\ie Euclidean), the geodesic path, which linearly interpolates ESRVFs, also linearly interpolates both the skeletal geometry and the associated branch thickness values. For visualization, we can map $g\optimal(t)$ back to the non-linear space of 3D tree-shaped structures $\preshapespaceThreeDTreees$ using the inverse ESRVF mapping, which has a closed analytical form; see~\cite{srivastava2010shape}.

\section{Shape space of 4D trees-like shapes}
\label{sec:4D_tree_shape_space}
With this representation, and after spatial registration, a 4D tree-like shape $\fourDTree$ becomes a time-parameterized  curve  $\fourDcurve$ in  $\preshapesrvts$, \ie $\fourDcurve: [0, 1] \to \preshapesrvts$ such that $\fourDcurve(t) \in \preshapesrvts $ is the ESRVFT of the 3D tree $\fourDTree(t)$. Thus, analyzing 4D tree-like shapes becomes a problem of analyzing these curves. This requires defining a shape space $\preshapespaceFourDcurve$ of such curves and equipping it with a proper Riemannian metric (Section~\ref{sect:4D tree shape space}). These will then be used to temporally register, compare, and compute geodesics between such curves (Section~\ref{sect:temopral_reg}). 

\subsection{The curve space of 4D tree-shaped structures}
\label{sect:4D tree shape space}

Let $\fourDcurve: [0, 1] \to \preshapesrvts$ be a curve in $\preshapesrvts$ representing the ESRVFT of a 4D tree-shaped object. Let $\preshapespaceFourDcurve$ be the space of all such curves. Since $\preshapespaceFourDcurve$ is of very high dimension, we propose to learn a low-dimensional subspace $\pcaspace$ of 3D trees, using Principal Component Analysis (PCA) on $\preshapespace_\srvfthreeDTree$. This way, 4D tree-like shapes become trajectories in this low-dimensional space. PCA assumes that the data follows a Gaussian distribution. However, when dealing with 3D tree-like shapes from different species, the distribution can be multi-modal. Thus, the subspace learned using PCA can be inaccurate. To address this issue, we propose to apply to the 3D trees in $\preshapespace_\srvfthreeDTree$ the Yeo-Johnson power transformation~\cite{weisberg2001yeo}, which transforms the data to follow a Gaussian distribution. We then apply PCA to the transformed data to learn the low-dimensional subspace $\pcaspace$ of the input 3D trees.

Let $\{\threeDTree_i\}_{i =1}^{\totalthreeDtree}$ be the set of all tree-like 3D objects under consideration and  $\{\srvfthreeDTree_i\}_{i =1}^{\totalthreeDtree}$ their corresponding ESRVFTs. We first compute the mean ESRVFT  $\meansrvft$ where $\meansrvft $ is the point in $\preshapesrvts$ that is as close as possible to all tree shapes in $\{\srvfthreeDTree_i\}_{i =1}^{\totalthreeDtree}$. The closeness is measured using the metric of Eqn.~\ref{eq:invariantdistance}:
\begin{equation}
\small{
	\meansrvft = \argmin_{(\rotation_i, \diffeotree_i, \permutetree_i)_{ i=1}^\totalthreeDtree}
			\sum_{i=1}^{\totalthreeDtree}d_{\preshapesrvts}(\srvfthreeDTree, (\srvfthreeDTree_i, \rotation_i, \reparm_i, \permutetree_i)).
}
\label{eq:karcher_mean}
\end{equation}

\noi This is the Karcher mean, which we compute  iteratively as follows;
\begin{enumerate}
	\item Set $\meansrvft =  \srvfthreeDTree_1$. 
	\item \label{step:for}For $i=2:\totalthreeDtree$,
		\begin{itemize}
			\item 
               Find $(\tilde\rotation_i,   \tilde\diffeotree_i, \tilde\permutetree_i)$ that optimally register  $\srvfthreeDTree_i$ onto $\meansrvft$ (Eqn.~\ref{eq:registration}). 
			\item \label{set:assign} Update $\meansrvft =  \frac{1}{2}(\meansrvft+(\srvfthreeDTree_i, \tilde\rotation_i, \tilde\diffeotree_i, \tilde\permutetree_i))$.
        \end{itemize}
	
	\item return $\meansrvft$ and $(\tilde\rotation_i,   \tilde\diffeotree_i, \tilde\permutetree_i)_{i=1}^{\totalthreeDtree}$. 
\end{enumerate}

\begin{figure}[tb]
     \centering
     \includegraphics[width=\linewidth]{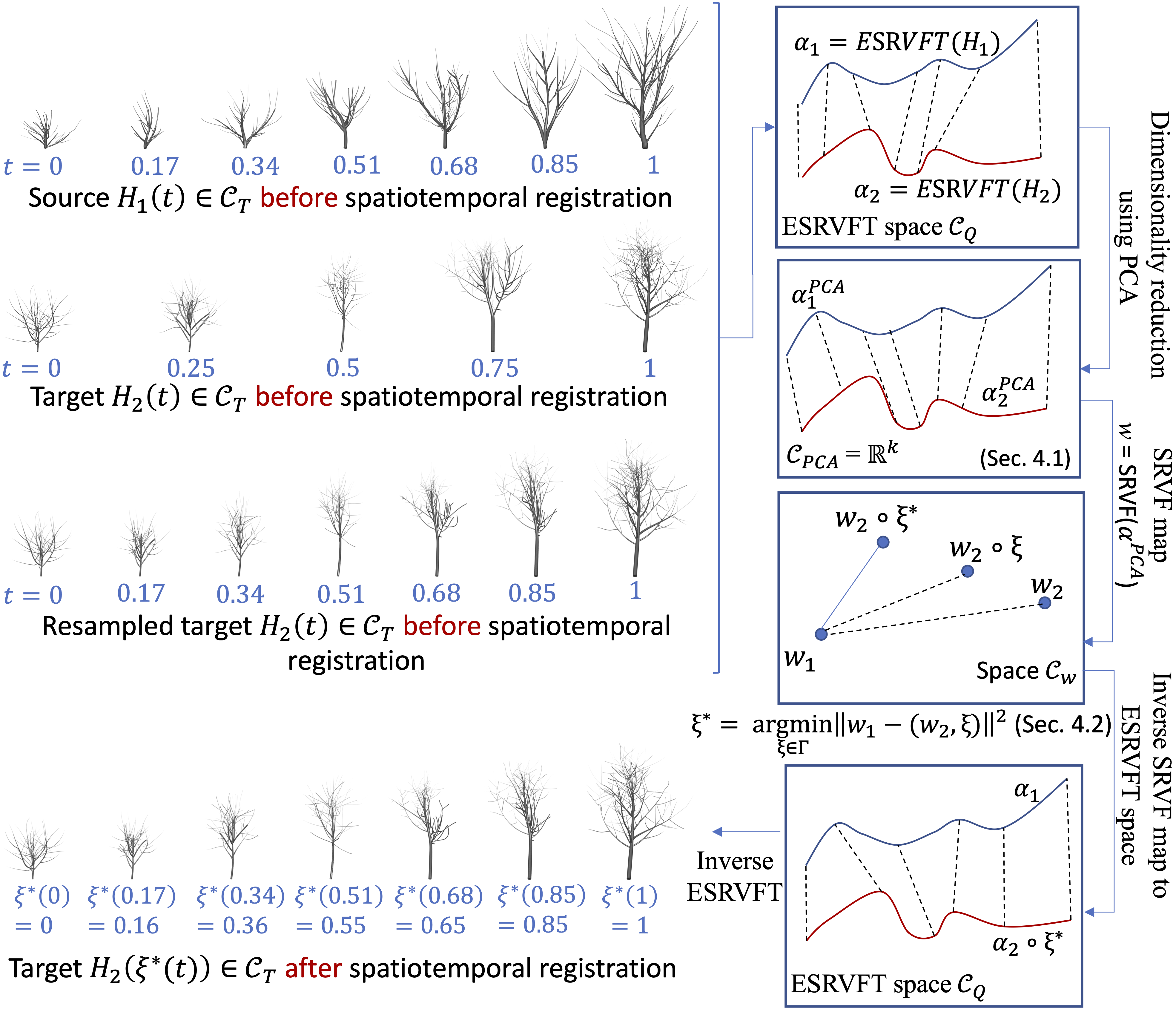} 
     \caption{Overview of the proposed framework for analyzing tree-like 4D objects. The core idea is to represent tree-like 4D objects $\fourDTree_1(t)$ and $\fourDTree_2(t)$ as curves in  $\preshapespaceThreeDTreees$, that are mapped to the ESRVFT space $\preshapesrvts$ which is infinite-dimensional but Euclidean. A PCA subspace $\pcaspace\subset \real^k$ is then learned to reduce dimensionality, allowing 4D shapes to be represented as finite-dimensional trajectories. To efficiently model temporal variability, these curves are further mapped to the SRVF space $\srvfpcaspace$, where the full elastic metric~\cite{srivastava2010shape} becomes the standard $\ltwo$ metric. We perform temporal alignment in this space by solving for the optimal reparameterization $\reparmcurve\optimal$ that minimizes $||\pcasrvf_1-(\pcasrvf_2,\reparmcurve)||$. After alignment, the registered curve is mapped back through the inverse SRVF and ESRVFT transformations to the original tree space $\preshapespaceThreeDTreees$ for visualization. This pipeline enables accurate spatiotemporal registration and statistical analysis of tree-like 4D objects.
     }
     \label{fig:all_mapping}
 \end{figure}

\noi One can map $\meansrvft$ 
 back to the original tree shape to obtain the mean tree $\meantree$, which can be used for visualization. Let $\srvfthreeDTree_i = (\srvfthreeDTree_i,  \tilde\rotation_i, \tilde\diffeotree_i, \tilde\permutetree_i)$ be the ESRVFT representation of $\srvfthreeDTree_i$ but optimally registered onto  the mean $\meansrvft$, and  $\bm{v}_i = \srvfthreeDTree_i - \meansrvft$.  
The  leading eigenvectors  $\Lambda_i$  of the  covariance matrix   $\displaystyle \covMatrix = \frac{1}{\totalthreeDtree-1}\sum_{i=1}^\totalthreeDtree\bm{v}_i\bm{v}_i^\top$ define the principal directions of variations while the corresponding eigenvalues $\lambda_i$ define the variance along the $i-$th eigenvector.  Thus, each ESRVFT  $\srvfthreeDTree$ can be modeled as a  linear combination of the $k$ leading eigenvectors: $
	\srvfthreeDTree = \meansrvft + \sum_{i=1}^k a_i\sqrt{\lambda_i} \Lambda_i.
$
The coefficients $a_i\in \real$ are obtained by projecting $\srvfthreeDTree$ onto the  eigenvectors $\Lambda_i$.   Thus, every tree shape $\srvfthreeDTree $ in the ESRVFT space can now be represented as a point $p = (a_1, \dots, a_k)^\top$ in $\pcaspace = \real^k$, which is Euclidean and has a finite dimension.

With this representation, a 4D tree-shape becomes a trajectory in $\pcaspace$, instead of the original ESRVFT space $\preshapespace_\srvfthreeDTree$.

\subsection{Execution rate-invariant metric}
\label{sec:rate_invariant_metric}

To compare, align, and summarize  4D trees defined as curves in the $\pcaspace$  space,  we require a proper metric that is invariant to the execution rate of the curves. Mathematically, variations in the execution rate correspond to time-warping and thus can be represented using orientation-preserving diffeomorphisms  $\reparmcurve: [0, 1] \to [0, 1] $ that map the temporal domain to itself. $\reparmcurve$ is said to be an orientation-preserving diffeomorphism if $\forall t_1, t_2 \in [0, 1], \text{ if } t_1 < t_2 \text{ then } \reparmcurve(t_1)  < \reparmcurve(t_2)$.
The temporal registration of two 4D   trees  $\fourDcurveinpcaone$ and $\fourDcurveinpcatwo \in \pcaspace$ then becomes the problem of finding the optimal time warping $\reparmcurve\optimal$ that brings the two curves as close as possible to each other, \ie
$
    \reparmcurve\optimal = \argmin_{\reparmcurve\in\reparmcurvespace} d(\fourDcurveinpcaone, \fourDcurveinpcatwo \circ \reparmcurve),
$. Here, $\reparmcurvespace$ is the space of all orientation-preserving diffeomorphisms $\reparmcurve: [0, 1] \to [0, 1] $ and $d(\cdot, \cdot)$ the metric.

We are left with the problem of defining the metric $d(\cdot, \cdot)$. As time-warping of trajectories corresponds to elastic deformation of curves, we follow the same approach as the one used to compare branches:  instead of using a complex non-linear elastic metric to measure the dissimilarity between two trajectories $\fourDcurveinpcaone$ and $\fourDcurveinpcatwo$, we first map them to their SRVF space using 
\begin{equation}
\label{eq:pcatosrvf}
\small{
    \pcasrvf(t)= \frac{\diffalpha}{\sqrt{\norm{\diffalpha}}}, \text{ if } \norm{\diffalpha}\neq 0, \text{ and } 0 \text{ otherwise.}
    }
\end{equation}

\noi Here,  $\diffalpha$ is the tangent vector to $\fourDcurveinpca$ at time $t$.  Working in the SRVF space has many benefits. \textbf{First}, if  $\srvffourDTree$ is the SRVF of a curve $\fourDcurveinpca$ then the SRVF of $\fourDcurveinpca \circ \reparmcurve$ is defined as $ (\srvffourDTree\circ \reparmcurve) \sqrt{\dot\reparmcurve}\|$, which we denote by $(\srvffourDTree,  \reparmcurve)$. \textbf{Second}, the elastic metric $d$ between two curves $\fourDcurveinpca_1$ and $\fourDcurveinpca_2$ in the original space reduces to the $\ltwo$ metric between their SRVFs, \ie 
\begin{equation} 
   d(\fourDcurveinpca_1, \fourDcurveinpca_2) = \norm{\srvffourDTree_1 - \srvffourDTree_2 }^2.    \label{eq:srvf_distance_trajectories_registration}
\end{equation}

\noi \textbf{Third}, under the $\ltwo$ metric, the action of the diffeomorphism  group $\reparmcurvespace$ is by isometries, \ie $\|\srvffourDTree_1, \srvffourDTree_2\| = \|(\srvffourDTree_1,  \reparmcurve), (\srvffourDTree_2, \reparmcurve)\|$. \textbf{Fourth}, the SRVF is invertible, up to translation. In other words, one can perform all the analysis tasks in the SRVF space $\srvfpcaspace$, which has an $\ltwo$ structure, and then map the results back to the original space for visualization without losing information. 

With this representation, a 4D tree-shaped object $\fourDcurve: [0, 1] \to \preshapesrvts$ becomes a trajectory  $\pcasrvf$ 
in the SRVF space $\srvfpcaspace$, which has an $\ltwo$ structure. Thus,  all the analysis tasks can be performed in this space using the standard $\ltwo$ metric. The results can then be mapped back to the original space for visualization.

\subsection{Temporal registration and geodesics}
\label{sect:temopral_reg}

Fig.~\ref{fig:all_mapping} summarizes the proposed temporal registration and geodesic computation process. To temporally register two curves $\srvffourDTree_1$ and $\srvffourDTree_2$, we find an optimal diffeomorphism $\reparmcurve\optimal: [0, 1] \to [0, 1] $ such that: 
\begin{equation}
    \label{eq:temp_reg}
    \reparmcurve\optimal=\underset{\reparmcurve\in\reparmcurvespace}{\argmin}\left\|\srvffourDTree_1-(\srvffourDTree_2, \reparmcurve)\right\|^2.
\end{equation} 

\noi We solve this optimization problem using dynamic programming introduced in~\cite{srivastava2010shape} while enforcing $\reparmcurve$ to be an orientation-preserving diffeomorphism, \ie  $\forall t_1, t_2, \text{ if } t_1 < t_2 \text{ then } \reparmcurve(t_1) = u_1 < \reparmcurve(t_2) = u_2$. 
With this formulation, the full process for the spatiotemporal registration can be summarized as follows; Let $V =\{\fourDTree_i\}_{i=1}^{\totalfourDtree}$ be a set of $\totalfourDtree$ 4D tree-shaped objects where $\fourDTree_i(t) \in \preshapespaceThreeDTreees$, for $t \in [0, 1]$, is a 3D tree-shape after normalization for translation.

\vspace{4pt}
\noi \textbf{Step 1: Spatial registration (Section~\ref{sect:srvf_3Dtree}).}
\begin{itemize}
    
        \item Map every 4D tree $\fourDTree_i$ to its ESRVFT representation $\fourDcurve_i$. Thus, we obtain a new set $\{\fourDcurve_i\}_{i=1}^{\totalfourDtree}$ such that $\fourDcurve_i(t) = \text{ESRVFT}(\fourDTree_i(t))$. Note that  $\fourDcurve(t)$ is a 3D tree $Q$ in the ESRVFT space of trees $\preshapesrvts$ (see Section~\ref{sect:srvf_3Dtree}). 
        
        \item Within-sequence  registration (Section~\ref{sect:spatial_registraion_within}):
            \begin{itemize}
                \item For $i=1$ to $\totalfourDtree$, spatially register every 3D tree-shape $\fourDcurve_i(t)$ in the $i-$th sequence $\fourDcurve_i$ to its next 3D tree-shape $\fourDcurve_i(t+1)$ in the sequence. 
                \item For simplicity of notation, let from now on  $\{\fourDcurve_i\}$ denote the new set. 
            \end{itemize}

        \item Cross sequence spatial registration of  $\fourDcurve_1$ onto $\fourDcurve_2$,
            \begin{itemize}
                \item Map $\fourDcurve_i, i \in \{1, 2\}$ to the PCA space $\pcaspace$  to  obtain   $\{ \fourDcurveinpca_i \}$.

                \item Interpolate, linearly, the samples of $\fourDcurveinpca_1$ and $\fourDcurveinpca_2$ and then resample them at equidistant time points.

                \item $\forall t$, spatially register  $\fourDcurveinpca_1(t)$ onto  $\fourDcurveinpca_2(t)$ (Section~\ref{sect:spatial_registraion_within}).
            \end{itemize}
    \end{itemize}

\vspace{4pt}
\noi\textbf{Step 2: Temporal registration (Section~\ref{sect:temopral_reg}).} 
    \begin{itemize}
        \item Map the spatially registered  $\fourDcurveinpca_i$  to their  SRVF representation $\pcasrvf_i$. 
        \item For  any two curves $\pcasrvf_1$ and $\pcasrvf_2$:
        \begin{itemize}
            \item Find $\reparmcurve\optimal$ that optimally aligns $\pcasrvf_2$ onto $\pcasrvf_1$\\ by solving Eqn.~\ref{eq:temp_reg}.
            
            \item Set $\Tilde{\pcasrvf}_2 \leftarrow  (\pcasrvf_2,\reparmcurve\optimal)$ and map it back to  $\preshapespaceThreeDTreees$  for visualization.
        \end{itemize}       
    \end{itemize}

\noi
\textbf{4D geodesics:} Since the $\ltwo$ metric in the SRVF space of curves $\srvfpcaspace$  is equivalent to the elastic metric,   the straight line  $\Lambda_\pcasrvf$ between $\pcasrvf_1$ and $\Tilde{\pcasrvf}_2$, given by $
    \Lambda_\pcasrvf(\tau)=(1-\tau)\pcasrvf_1+\tau\Tilde{\pcasrvf}_2,\ \tau\in[0,1]
$, is equivalent to the geodesic path between their corresponding 4D trees. Here, $\Lambda_\pcasrvf(0)= \pcasrvf_1$ and $\Lambda_\pcasrvf(1)= \Tilde{\pcasrvf}_2$. To visualize the geodesics, we map the curve $\Lambda_\pcasrvf(\tau)$ to the original space of trees as follows:
\begin{equation}
\label{eq:v_pca}
    \pcasrvf \xrightarrow{SRVF \ inversion} \fourDcurveinpca,
\nonumber
\end{equation}
\begin{equation}
\label{eq:pca_alpha}
    \fourDcurveinpca \xrightarrow{Inverse \ PCA} \fourDcurve,
\nonumber
\end{equation}
\begin{equation}
\small{
\begin{aligned}
  \label{eq:alpha_q}
   \fourDcurve=(\fourDcurve_1,\dots, \fourDcurve_d)={\{\srvfthreeDTree^1,\dots, \srvfthreeDTree^d\}}, 
\nonumber  
\end{aligned}}
\end{equation}
\begin{equation}
\small{
\begin{aligned}
    \label{eq:q_org}
    {\{\srvfthreeDTree^1,\dots, \srvfthreeDTree^d\}} \xrightarrow{ESRVF \ inversion \ } {\{\threeDTree^1,\dots, \threeDTree^d\}} = \fourDTree,
    \nonumber
\end{aligned}
}
\end{equation}

\noi Here,  $d$ is the total number of points sampled along a curve. 

Fig.~\ref{fig:temporal_Reg_different_target}  shows an example of temporal registration while 
Fig.~\ref{fig:geod_after_reg1} shows an example of a geodesic path between two 4D trees $\pcasrvf_1$ and $\Tilde{\pcasrvf}_2$, where each row corresponds to one 4D tree. The first row corresponds $\Lambda_\pcasrvf(0)= \pcasrvf_1$ and the last row corresponds $\Lambda_\pcasrvf(1)= \pcasrvf_2$. The middle row represents the mean 4D tree shape between $\pcasrvf_1$ and $\Tilde{\pcasrvf}_2$,  which corresponds to  $\Lambda_\pcasrvf(0.5)$.

\section{Statistics on 4D tree shapes}
\label{sec:statistical_analysis}
In this section, we detail how these representations and computational tools can be used to compute a 4D atlas of tree-like shapes with full geometry. This includes computing the statistical mean and modes of variation of a collection of tree-shaped 4D structures. 

Let $\{\fourDTree_1, \dots, \fourDTree_\totalfourDtree\}$ denote  a collection of $\totalfourDtree$ 4D tree shapes and $\{\pcasrvf_1, \dots, \pcasrvf_\totalfourDtree\}$  be their corresponding curves in $\srvfpcaspace$. Since the space $\srvfpcaspace$ is Euclidean, we can compute all the statistics in $\srvfpcaspace$  using standard tools and then map the results back to the original space for visualization. 
We start by spatially registering all the input 4D tree shapes onto each other using the procedure described in Sections~\ref{sect:spatial_registraion_within} and~\ref{sect:temopral_reg}. For simplicity of notation, we let $\{\pcasrvf_1, \dots, \pcasrvf_\totalfourDtree\}$ be the spatially registered 4D trees. Next, we use the tools developed in this paper to compute the mean and modes of variation, and build generative models that enable the synthesis of new  4D tree-like shapes.

\vspace{3pt}
\noi\textbf{Mean of 4D tree-like shapes.} The statistical mean of a set of 4D trees is the 4D tree that is the closest to all samples in the dataset. This is referred to as the  Karcher mean $\Bar{\pcasrvf}$, which can be computed using the following optimization procedure: 
\begin{enumerate}
	\item Set $\Bar{\pcasrvf} =  \pcasrvf_1$. 
	\item \label{step:for2}For $i=1:\totalfourDtree$
		\begin{itemize}
			\item Temporally register  $\pcasrvf_i$ onto $\Bar{\pcasrvf}$ using Step 2 of the Algorithm of Section~\ref{sect:temopral_reg}, to obtain  $\Tilde{\pcasrvf_i}$. 
			
                \item Let $\reparmcurve_i$ be the reparameterization that temporally aligns $\pcasrvf_i$ onto $\Bar{\pcasrvf}$.
            \item Update $\Bar{\pcasrvf} =  \frac{1}{i}(\Bar{\pcasrvf}+\Tilde{\pcasrvf_i}) $.
		\end{itemize}	
	\item return $\Bar{\pcasrvf}$ and   $\{\reparmcurve_i, \Tilde{\pcasrvf}_i \}_{i=1}^ \totalfourDtree$. 
\end{enumerate}

\noi The  mean curve $\Bar{\pcasrvf}$ can be mapped back, using the procedure presented in Section~\ref{sect:temopral_reg},  to a 4D tree, $\Bar{\fourDTree}$, which  represents the mean of the 4D tree-shaped structures in the dataset.

\vspace{6pt}
\noindent
\textbf{Modes of variation.} Since the SRVF space of curves $\srvfpcaspace$ is Euclidean, we compute the principal directions of variation using  PCA. Since the curves $\{\Tilde{\pcasrvf}_i\}_{i=1}^{\totalfourDtree}$ are now registered and their mean $\Bar{\pcasrvf} \in \srvfpcaspace$ computed, we perform PCA  by  computing the covariance matrix  $\covMatrix=\frac{1}{\totalfourDtree-1}\sum_{i=1}^\totalfourDtree(\Tilde{\pcasrvf}^i-\Bar{\pcasrvf})(\Tilde{\pcasrvf}^i-\Bar{\pcasrvf})^{\top}$. 
Let $\{e_i\}_{i=1}^k$ be the leading eigenvectors of the covariance matrix $\covMatrix$  and $\{\delta_i\}_{i=1}^k$ the corresponding eigenvalues. The eigenvectors represent the principal directions of variation of the curves. The projection of any curve, and thus a 4D tree, onto the $i-$th principal direction is given by $\pcasrvf_\tau=\Bar{\pcasrvf}+\tau\sqrt{e_i}\delta_i,\ \tau\in\real$. For visualization, we map $\pcasrvf_\tau$ back to the original space of 4D trees using the inversion procedure described in Section~\ref{sect:temopral_reg}. 

\vspace{3pt}
\noindent
\textbf{4D tree-like shape synthesis:} Given the mean curve $\Bar{\pcasrvf}$ and the $k$ principal directions of variation, $e_i, i=1,\dots, k$, one can generate  a 4D tree-like shape by  sampling $k$ real values $a_i\in\real$ and then computing a new shape as:
\begin{equation}
     \pcasrvf= \Bar{\pcasrvf} + \sum_{i=1}^k a_i\sqrt{\delta}_i e_i, \ a_i\in\real.
     \label{eq:generation}
\end{equation} 

\noindent The generated curve $\pcasrvf$ can be mapped back to the original tree-shape space for visualization using the inversion procedure discussed in Section~\ref{sect:temopral_reg}. Note that one can restrict the generation process to plausible 4D shapes only by enforcing the weights $a_i$ to be within a specified range. This ensures that the generated 4D tree-like shapes remain in the vicinity of the mean shape.

\section{Results and discussion}
\label{sec:result}
We evaluate the proposed framework using synthetic 4D tree models from Globe plants\footnote{\url{https://globeplants.com}} and real 4D plants from Pheno4D dataset~\cite{pheno4d}. The models were spatially and temporally unregistered; see the supplementary material for more details. All the codes were implemented using  MATLAB(2024), and the experiments were run on a $2.00$GHz Intel Core i$9$ processor with $128$GB of RAM.

\begin{figure}[tb]
    \centering
    \includegraphics[width=\linewidth]{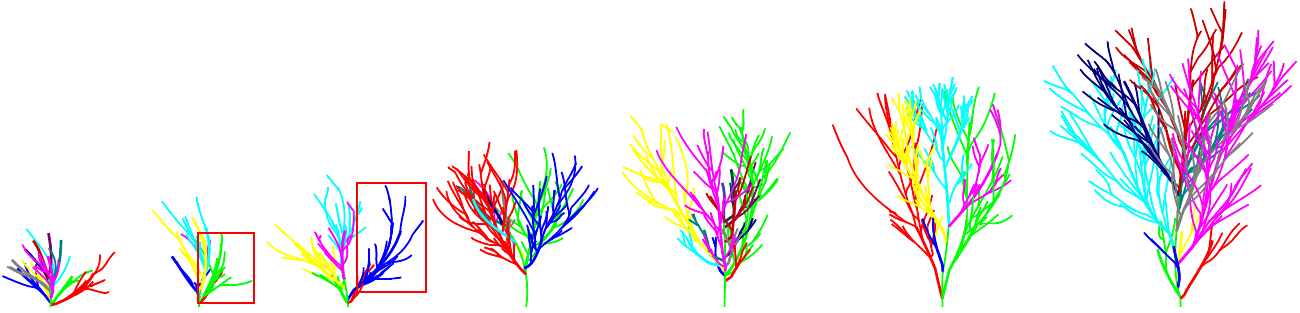}
    \small{\textbf{(a)} A 4D tree before spatial registration.}
    \includegraphics[width=\linewidth]{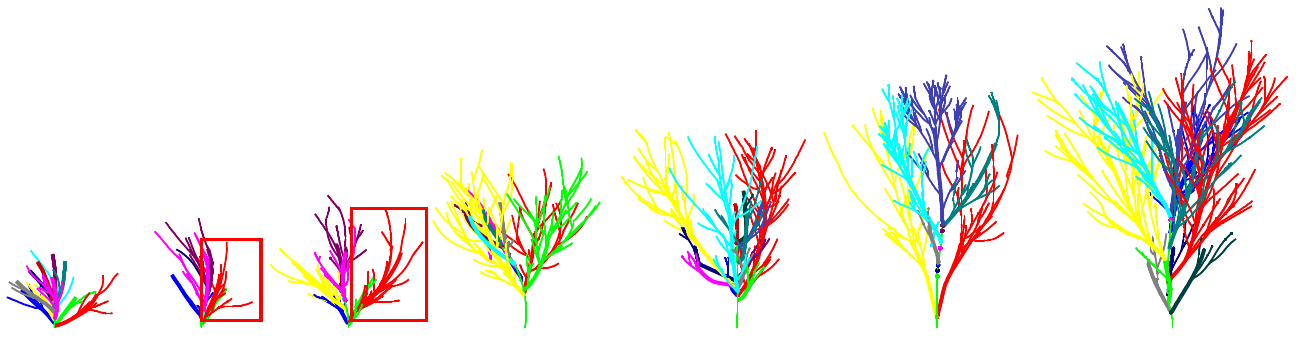}
    \small{\textbf{(b)} The same 4D tree after spatial registration.}
    
    \caption{Visual example of spatial registration applied to a 4D tree-like shape using the proposed framework. \textbf{(a)} The input 4D tree before registration, showing branch misalignments over time. \textbf{(b)} The same tree after spatial registration, with improved alignment of branch geometry and structure. Color coding indicates corresponding branches up to the second hierarchical layer, highlighting the consistency achieved through registration. Fig.~\ref{fig:zoom_spatial_registration_Level2} provides a zoomed-in view of the third-level branch correspondences for the regions highlighted with red boxes.}
    \label{fig:spatial_registration_L4}
\end{figure}

\begin{figure}[tb]
    \centering
    \begin{tabular}{@{}c@{}c@{}}
        \includegraphics[width=0.5\linewidth]{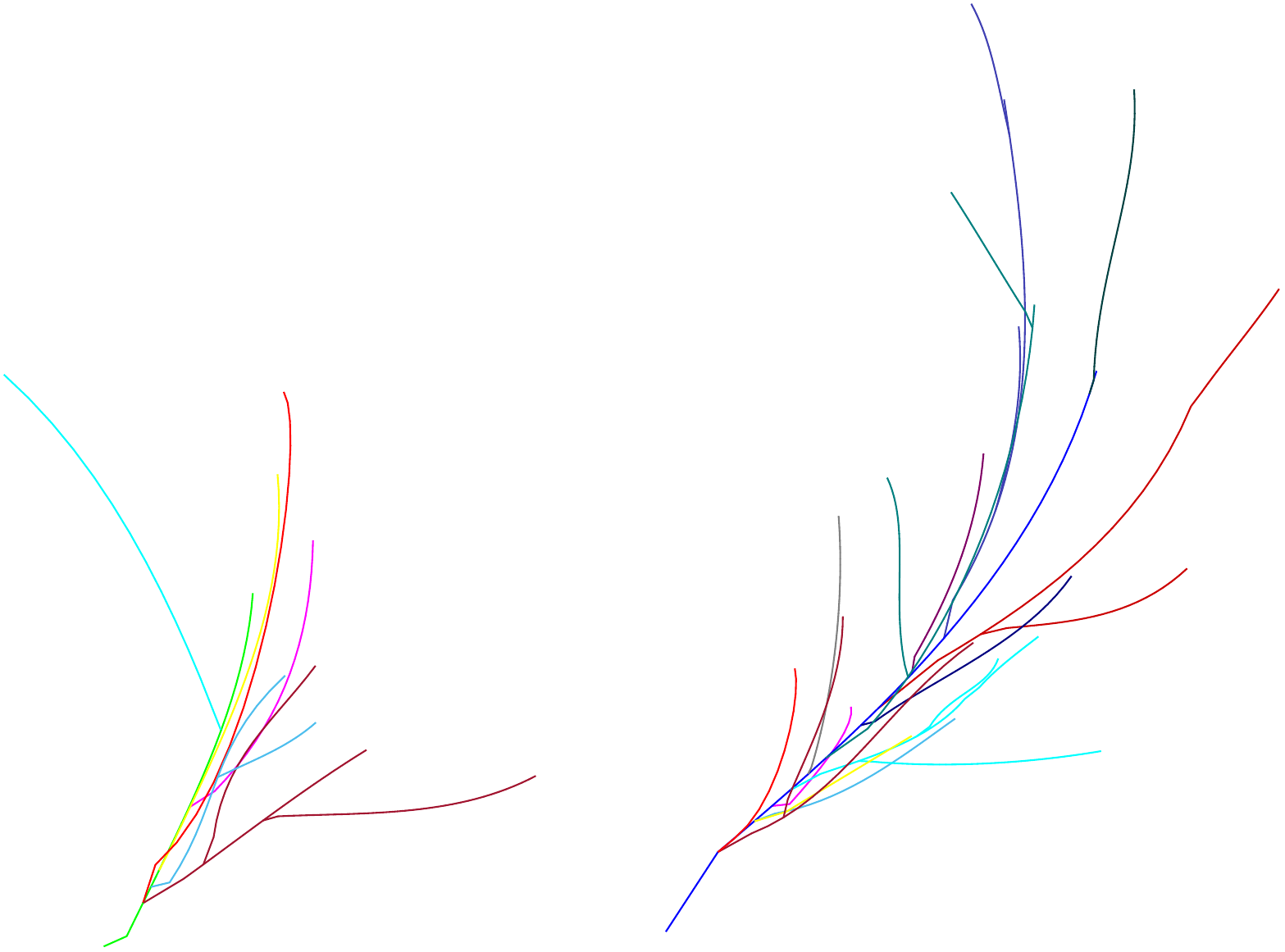} &
        \includegraphics[width=0.5\linewidth]{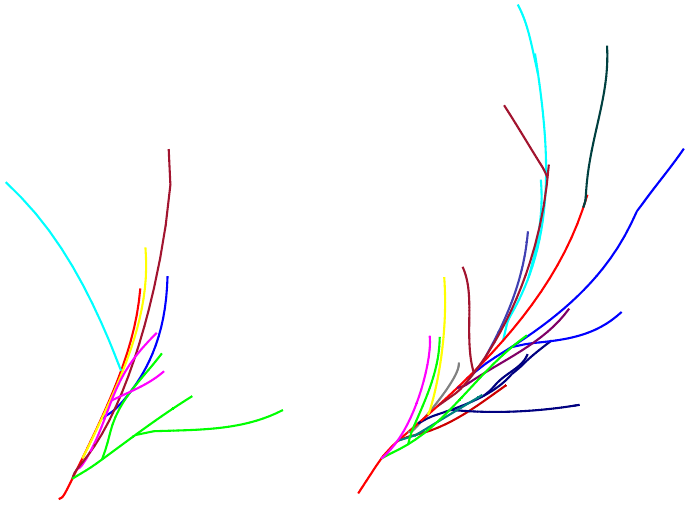} \\
        
        \small{\textbf{(a)} Before spatial registration.} &        
        \small{\textbf{(b)} After spatial registration.}
    \end{tabular}
    \caption{Zoomed-in visualization of third-level branch correspondences for the region highlighted with a red box in Fig~\ref{fig:spatial_registration_L4}, shown (a) before and (b) after spatial registration. The improved alignment of finer branch structures demonstrates the effectiveness of the proposed registration method at deeper hierarchical levels.}
    \label{fig:zoom_spatial_registration_Level2}
\end{figure}

\begin{figure}[tb]
    \centering
    \includegraphics[width=\linewidth]{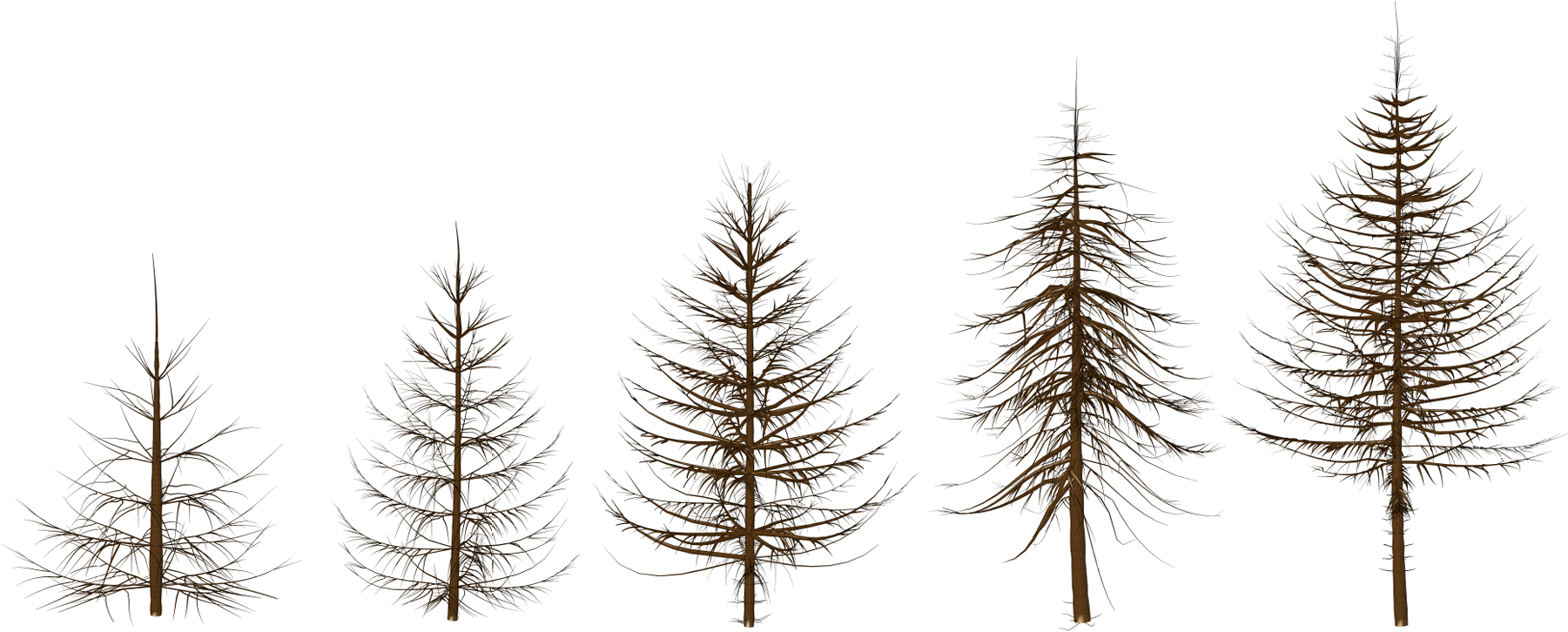}
    \small{\textbf{(a)} A complex 4D tree with a large number of side branches.}
    \includegraphics[width=\linewidth]{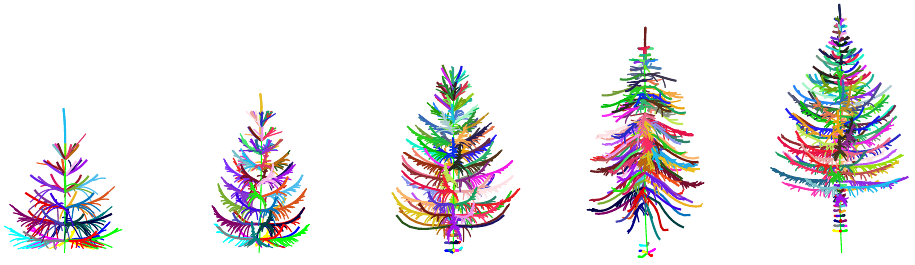}
    
    \small{\textbf{(b)} Before spatial registration.
    }
    \includegraphics[width=\linewidth]{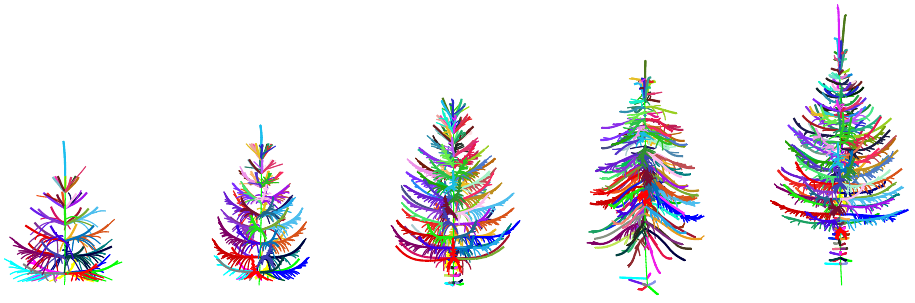}
    \small{\textbf{(c)} After spatial registration.
    }
    \caption{Spatial registration results for a complex 4D tree with a dense branching structure, \textbf{(a)} shows the raw input geometry, \textbf{(b)} and \textbf{(c)} illustrate the tree before and after registration, respectively, with correspondences color-coded up to the second level of the branching hierarchy. The post-registration alignment highlights improved consistency across branch structures over time.  } 
    \label{fig:spatial_registration_large_number_of_branches}
\end{figure}

\subsection{Spatial registration} 
Fig.~\ref{fig:spatial_registration_L4} (Four-layer structure)
and Fig. 1 (Two-layer structure) in the supplementary material show an example of the spatial registration of the 3D trees that belong to the same growing 4D tree model.  For clarity, we color-code the correspondences only up to level two of the tree hierarchy. Fig.~\ref{fig:zoom_spatial_registration_Level2} shows a zoom-in of the third level correspondence on the branches highlighted with a red box in Fig.~\ref{fig:spatial_registration_L4}. Visually, we can see that our framework produces plausible spatial registrations. 
Fig.~\ref{fig:spatial_registration_large_number_of_branches} shows another example of registration between complex 3D trees that have a large number of side branches. Despite their high complexity, our approach is able to find correct branch-wise correspondences between the 3D trees. We quantitatively evaluate the quality of the proposed spatial registration method and compare it to state-of-the-art techniques such as~\cite{chebrolu2021registration} and~\cite{pan2021multi}. We use three evaluation metrics: \textbf{(1)} cycle consistency,  \textbf{(2)} description length, and \textbf{(3)} computation time. 

\begin{table*}
   \centering
    \caption{Evaluation of spatial registration based on cycle consistency errors $(\downarrow)$ for the Globe plants and Pheno4D datasets. We report the mean, median, and standard deviation of errors at different thresholds $\epsilon$. Our method significantly outperforms Pan \etal~\cite{pan2021multi}, demonstrating improved consistency across registration cycles. Lower percentages indicate better performance.} 
    \label{tab:cycle_consistancy}
   
    \begin{tabular}{|@{ }c@{ }|@{ }c@{ }|@{ }c@{ }|@{ }c@{ }|@{ }c@{ }|@{ }c@{ }|@{ }c@{ }|@{ }c@{ }|@{ }c@{ }|@{ }c@{ }|@{ }c@{ }|@{ }c@{ }|c@{ }|}
        \hline
        & \multicolumn{6}{c|}{\href{https://globeplants.com}{Globe plants}}  &  \multicolumn{6}{c|}{Pheno4D~\cite{pheno4d}}\\
        \cline{2-13}
             & \multicolumn{3}{c|}{\textbf{Our method}}  &  \multicolumn{3}{c|}{Pan \etal~\cite{pan2021multi}} & \multicolumn{3}{c|}{\textbf{Our method}} & \multicolumn{3}{c|}{Pan \etal~\cite{pan2021multi}}\\
        \cline{2-13}
          $\epsilon$   & Mean & Median & Std. & Mean & Median & Std. & Mean & Median & Std. & Mean & Median & Std.\\
        \hline
        $0.1$ & $\bf{0.08\%}$ & $\bf{0.09\%}$  & $\bf{0.08\%}$  & $20\%$ & $21\%$ & $9.6\%$ &  $\bf{0.94\%}$ & $\bf{0.08\%}$  & $\bf{2.01\%}$  & $10.33\%$ & $7.09\%$ & $10.60\%$ \\
        \hline
        $0.05$ & $\bf{0.17\%}$  & $\bf{0.2\%}$  &  $\bf{0.09\%}$ & $28.2\%$ & $28\%$ & $4.3\%$ & $\bf{2.00\%}$  & $\bf{0.15\%}$  &  $\bf{3.62\%}$ & $15.89\%$ & $17.88\%$ & $10.79\%$ \\  
        \hline
        $0.02$ & $\bf{1.8\%}$  &  $\bf{1.8\%}$ &  $\bf{0.89\%}$ & $32\%$ & $32.5\%$ & $1.3\%$ & $\bf{2.87\%}$  &  $\bf{0.65\%}$ &  $\bf{4.17\%}$ & $20.39\%$ & $22.22\%$ & $9.83\%$ \\
        \hline
        $0.01$ & $\bf{3.3\%}$ & $\bf{3.5\%}$ &  $\bf{1.7\%}$ & $33\%$ & $33\%$  &  $0.63\%$ & $\bf{5.45\%}$ & $\bf{2.76\%}$ &  $\bf{6.15\%}$ & $21.95\%$ & $22.22\%$  &  $8.64\%$\\
        \hline
    \end{tabular}
\end{table*}

\begin{figure}[tb!]
        \centering
        \includegraphics[trim={0 0 0 1cm}, clip, width=0.8\linewidth]{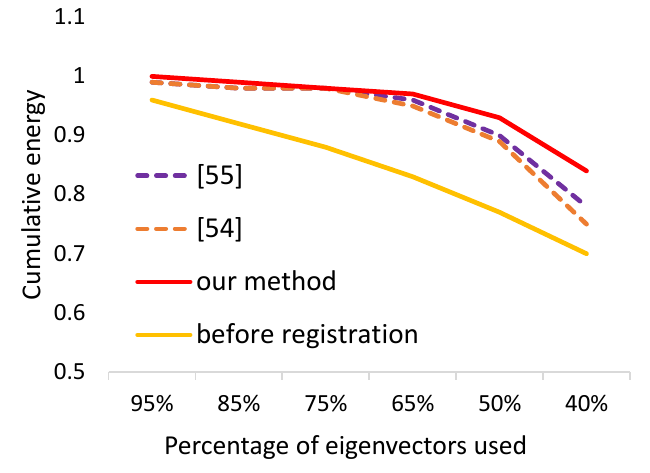}
        \caption{Cumulative energy against the percentage of eigenvectors used to capture shape variability, computed on Pheno4D~\cite{pheno4d} dataset. Previous methods exhibit a high loss of energy when using less than $65\%$ of the eigenvectors.}
        \label{fig:description_length}
\end{figure}

\vspace{4pt}
\noi \textbf{(1) Cycle consistency error}.
Given source and target 3D trees, the spatial registration process maps a point $\textbf{x}$ on the source to a point $\textbf{y}$ on the target. We then map $\textbf{y}$, using the same registration procedure, back onto the source tree to lead to a point $\textbf{x}'$. The registration procedure is accurate if  $\textbf{x}'$ and $\textbf{x}$ are very close to each other. Thus, we define the cycle consistency-based registration error as the percentage of points $\textbf{x}$ whose distance $\|\textbf{x} - \textbf{x}'\|$ is higher than a threshold $\epsilon$.  We vary $\epsilon$ between $0.01$ and $0.1$ and report in Table~\ref{tab:cycle_consistancy}  the mean, median, and standard deviation of this error for our method and~\cite{pan2021multi} (the lower the better). As we can see, our method significantly outperforms the state-of-the-art.

\vspace{4pt}
\noi\textbf{(2) Description length.} It is defined as the number of eigenvectors needed to describe $x\%$ of the variability within the dataset.  $x$ is referred to as the cumulative energy. From Fig.~\ref{fig:description_length}, we can see that our method shows stable preservation of energy compared to~\cite{pan2021multi} and~\cite{chebrolu2021registration}.

\vspace{4pt}
\noi\textbf{(3) Computation time.} On average, the computational time varies from $5$ s to $2.7$ hours for the 3D trees in the Globe plant dataset, whose number of branches ranges from $500$ to $15,000$. Pan \etal~\cite{pan2021multi}, on the other hand, require $3$ minutes to $6$ hours for the same trees. On the Pheno4D~\cite{pheno4d} dataset, on average, Chebrolu \etal~\cite{chebrolu2021registration} require $30$ s, Pan \etal~\cite{pan2021multi} require $20.1$ s, while our method is significantly faster as it
only requires  $2.3$ s on average. 

\begin{table}[t]
\centering
\caption{Quantitative evaluation of temporal registration errors $(\downarrow)$ between pairs of 4D tree-shaped objects. We report the mean, median, and standard deviation of the alignment error before and after temporal registration across three datasets: Globe plants, Pheno4D (Tomato), and Pheno4D (Maize). In subtable (a), the errors are computed between pairs of source and target 4D trees. In subtable (b), the evaluation is performed between a 4D tree and its randomly time-warped version. In both settings, the proposed method significantly reduces temporal misalignment, demonstrating its effectiveness in synchronizing growth trajectories across tree-like 4D structures.}
\label{tab:temporal_reg_error}

\begin{tabular}{|@{}c@{ }|@{ }c@{ }|@{ }c@{ }|@{ }c@{ }|@{ }c@{ }|@{ }c@{ }|@{ }c@{}|}
    \hline
    \textbf{Dataset} &\multicolumn{3}{|c|}{\textbf{Before reg.}} &
    \multicolumn{3}{c|}{\textbf{After reg.}} \\ 
    \cline{2-7}
    &{Mean} &
   {Med.}&
    {Std.} &
     {Mean} &
     {Med.} &
    Std. \\ 
    
    \hline
   \href{https://globeplants.com}{Globe plants}&\multicolumn{1}{|c|}{$159$} &
  \multicolumn{1}{c|}{$184$}&
  \multicolumn{1}{c|}{$67$} &
  \multicolumn{1}{c|}{$\textbf{67.7}$} &
  \multicolumn{1}{c|}{$\textbf{81}$} &
  \multicolumn{1}{c|}{ $\textbf{26.2}$}\\ 
  
    \hline
    Pheno4D~\cite{pheno4d} &
  $382$&
  $384$&
  $69$&
  $\textbf{279}$ &
  $\textbf{282}$ &
  $\textbf{47}$\\ 
  (Tomato plants) & &  & & & &\\
  
 \hline
 \multicolumn{1}{|r|}{Pheno4D~\cite{pheno4d}} &
  \multicolumn{1}{c|}{$59$} &
  \multicolumn{1}{c|}{$60$} &
  \multicolumn{1}{c|}{$12$} &
  \multicolumn{1}{c|}{$\textbf{34}$} &
  \multicolumn{1}{c|}{$\textbf{37}$} &
  $\textbf{11}$ \\
  \multicolumn{1}{|r|}{(Maize plants)} & \multicolumn{1}{c|}{}& \multicolumn{1}{c|}{}& \multicolumn{1}{c|}{}& \multicolumn{1}{c|}{}& &\\
  
  \hline
   \multicolumn{7}{c}{\textbf{(a)} Between a source and a target 4D tree.}\\ 
 \multicolumn{7}{c}{}\\
\hline
 
  \textbf{Dataset} &\multicolumn{3}{|c|}{\textbf{Before reg.}} &
  \multicolumn{3}{c|}{\textbf{After reg.}} \\ 
  \cline{2-7} 

  &{Mean} &
   {Med.}&
    {Std.} &
     {Mean} &
     {Med.} &
    Std. \\ 
  
  \hline
 \href{https://globeplants.com}{Globe plants}&\multicolumn{1}{|c|}{$77.8$} &
  \multicolumn{1}{c|}{$60$} &
  \multicolumn{1}{c|}{$43$} &
  \multicolumn{1}{c|}{$\textbf{14.8}$} &
  \multicolumn{1}{c|}{$\textbf{11}$} &
  $\textbf{7.9}$\\ 
  
  \hline
  \multicolumn{1}{|r|}{Pheno4D~\cite{pheno4d}} &
  \multicolumn{1}{c|}{$356$} &
  \multicolumn{1}{c|}{$375$} &
  \multicolumn{1}{c|}{$138$} &
  \multicolumn{1}{c|}{$\textbf{26}$} &
  \multicolumn{1}{c|}{$\textbf{27}$} &
  $\textbf{5}$\\ 

    \multicolumn{1}{|r|}{(Tomato plants)} & \multicolumn{1}{c|}{}& \multicolumn{1}{c|}{}& \multicolumn{1}{c|}{}& \multicolumn{1}{c|}{}&\multicolumn{1}{c|}{}&\\
  \hline
  \multicolumn{1}{|r|}{Pheno4D~\cite{pheno4d}} &
  \multicolumn{1}{c|}{$71$} &
  \multicolumn{1}{c|}{$70$} &
  \multicolumn{1}{c|}{$8$} &
  \multicolumn{1}{c|}{$\textbf{15}$} &
  \multicolumn{1}{c|}{$\textbf{16}$} &
  $\textbf{3}$ \\
     \multicolumn{1}{|r|}{(Maize plants)} & \multicolumn{1}{c|}{}& \multicolumn{1}{c|}{}& \multicolumn{1}{c|}{}& \multicolumn{1}{c|}{}&\multicolumn{1}{c|}{} &\\
  \hline
   \multicolumn{7}{c}{\textbf{(b)} Between a 4D tree and its randomly warped version.} \\  
\end{tabular}%
\end{table}

\begin{figure}[tb]
    \centering
    \includegraphics[width=\linewidth]{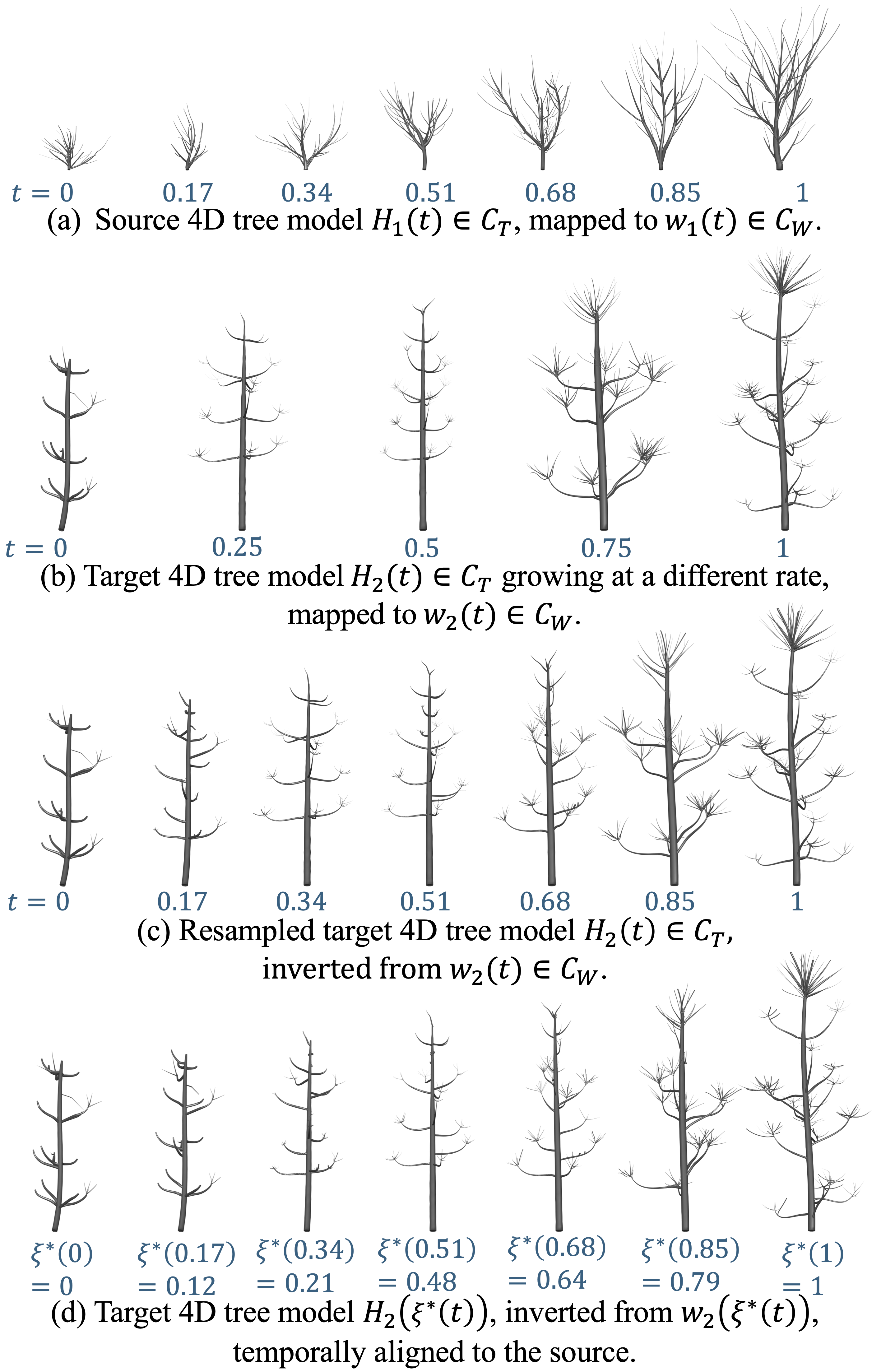}
    \caption{Illustration of temporal registration between two 4D trees, exhibiting different growth rates. \textbf{(a)} Source 4D tree. \textbf{(b)} Target 4D tree model exhibiting asynchronous growth relative to the source. Both source and target have been captured at different points in time $t \in [0, 1]$ and at different temporal resolutions.  \textbf{(c)} Uniformly resampled target to match the temporal resolution of the source. \textbf{(d)} Result after applying optimal reparameterization $\reparmcurve\optimal$ computed with our method, showing that the target is temporally aligned with the source.} 
    \label{fig:temporal_Reg_different_target}
\end{figure}

\subsection{Temporal registration}  
We evaluate the quality of temporal registration by measuring the geodesic distance between 4D tree shapes before and after registration (the smaller the better). We consider two cases. In the \textbf{first} case, we randomly choose $10$  pairs of 4D tree models. Each pair is composed of a source and a target 4D tree. Then, we compute the geodesic distance between the source and target before and after temporal registration. Table~\ref{tab:temporal_reg_error}(a) reports the mean, median, and standard deviation of those geodesic distances. Note that the residual error is primarily due to the differences in the branching structures. 

In the \textbf{second} case, we temporally warp, using random diffeomorphisms, ten 4D tree models,  register every warped 4D tree model to its original 4D tree model, and then measure the quality of the temporal registration using the geodesic distance between the registered 4D model and the original 4D tree model. As one can see from  Table~\ref{tab:temporal_reg_error}(b),  our proposed temporal registration can align two 4D tree shapes with minimal error.  Fig.~\ref{fig:all_mapping} shows a visual example of temporal registration between a source and a target 4D trees. Fig.~\ref{fig:temporal_Reg_different_target} shows another example before and after temporal registration. From both examples, we can see that the target is better aligned with the source after applying our algorithm. Fig. 3 in the supplementary material shows the temporal registration result of a target 4D tree that is a warped version of a source 4D tree. We can see that after temporal registration, the warped target becomes well synchronized with the source. 

\vspace{4pt}
\noi\textbf{Computation time.} Overall, the temporal registration of a pair of 4D trees takes $4$ to $30$ ms when performed in the proposed PCA space, whereas it takes $10$ to $20$ minutes when performed in the high-dimensional space of ESRVFTs. 

\begin{figure}[tb]
    \centering
    \includegraphics[width=.95\linewidth]{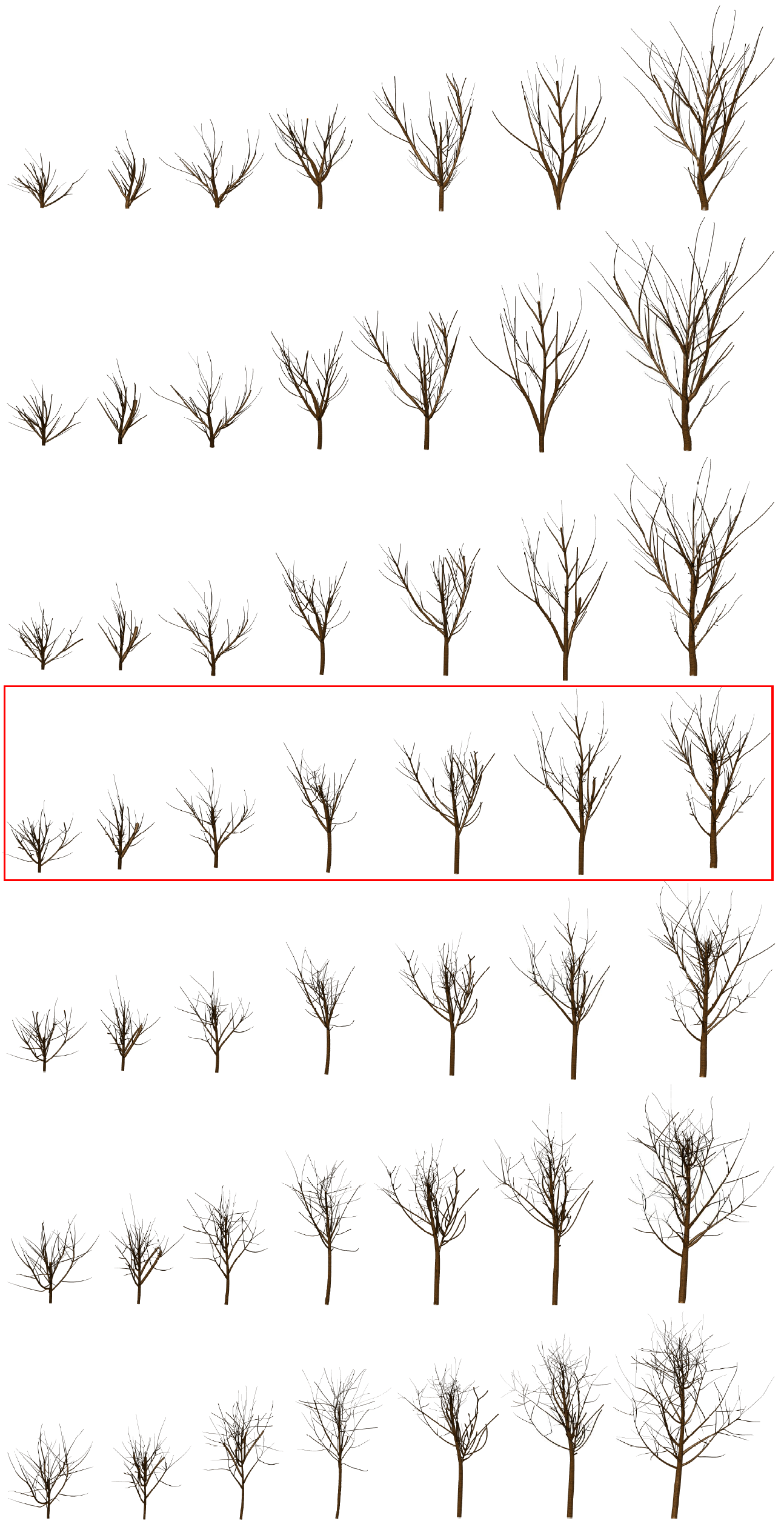}
    \caption{Visualization of a 4D geodesic path computed between a source 4D tree model (top row) and a target 4D tree model (bottom row), following spatiotemporal registration. The path is constructed in the SRVF space $\srvfpcaspace$, and mapped back to the tree domain. The highlighted middle row shows the intermediate interpolations along the geodesic, with the central frame representing the 4D mean shape. This illustrates smooth and structurally consistent deformation between two complex tree-like 4D objects. }
    \label{fig:geod_after_reg1}
\end{figure}

\subsection{Geodesics between 4D trees}
\label{sect:geodesic}
We compute the geodesic between pairs of 4D tree models. For each example, we show the geodesic before and after spatiotemporal registration.
Fig.~\ref{fig:geod_after_reg1} shows an example of a 4D geodesic after spatiotemporal registration, where Fig. 4 in the supplementary material shows the geodesic between the same pair, before spatiotemporal registration. The supplementary material shows more examples of 4D geodesics before and after spatiotemporal registration.
Note that in the former case, the source and target 4D trees have different execution rates and thus are not temporally aligned. In each figure, the top row corresponds to the source, the bottom row to the target, and the in-between rows correspond to 4D tree models sampled at equidistances along the geodesic path between the source and target. 
As we can see, after spatiotemporal registration, the geodesic path becomes more natural, whereas it suffers significant shrinkage before spatiotemporal alignment due to \textbf{(1)} the misalignment of long branches with either null or smaller branches, and \textbf{(2)} inappropriate pointwise correspondence between branches that leads to unnatural branch sliding and distorted geodesic paths. This shrinkage effect can also be observed between trees with the same number of branches after registration, when the analysis is performed in the original space of trees. Our proposed  ESRVF formulation addresses this issue. We discuss this case in Section~\ref{sect:ablation}.

\vspace{4pt}
\noi\textbf{Computation time.} On average, the geodesic computation between a pair of 4D tree models takes $20$ to $30$ ms when performed in the low dimensional PCA space.

\begin{figure}[tb]
    \centering
    \includegraphics[width=.95\linewidth]{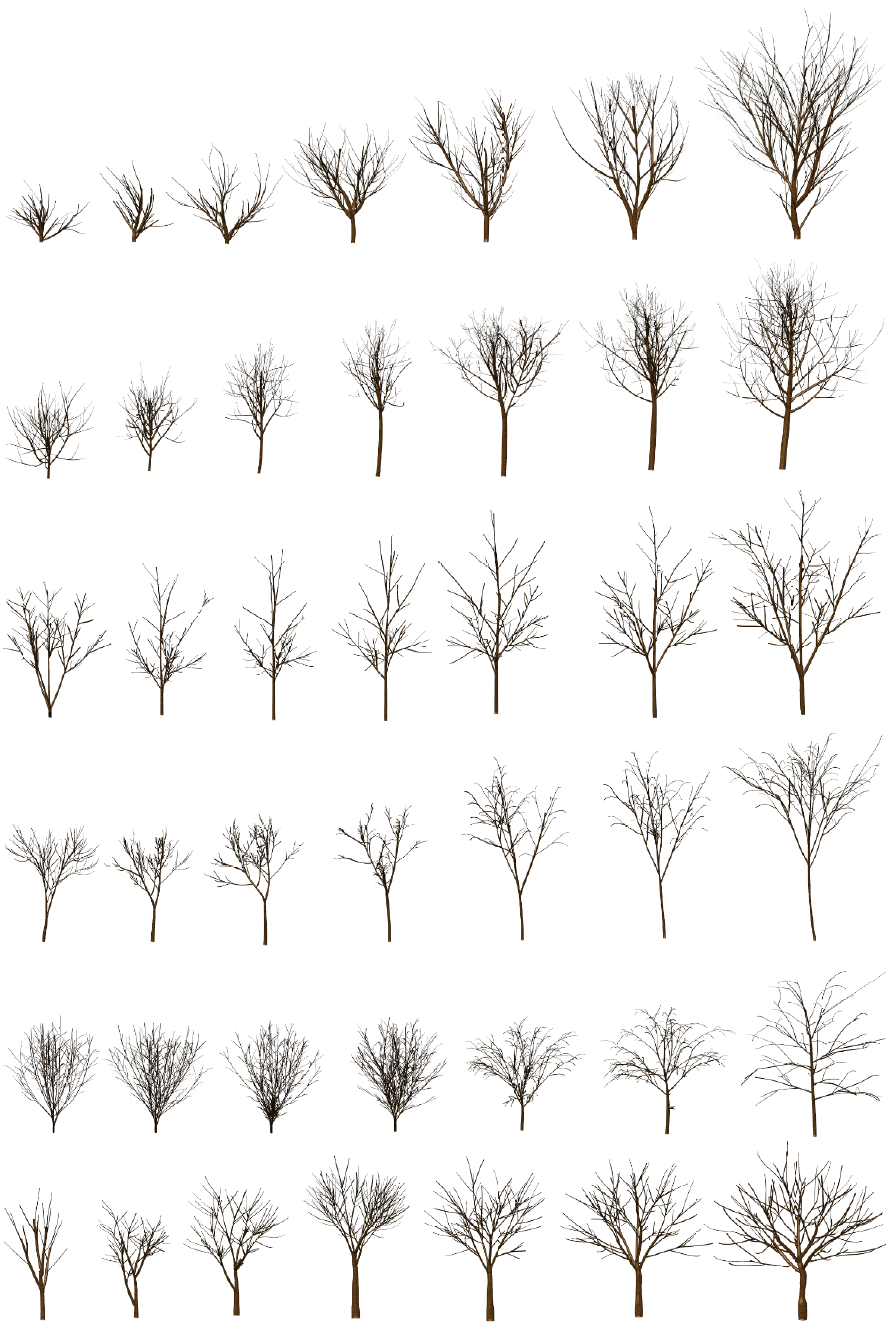}
    \caption{Visualization of the first set (Set 1) of 4D tree models after spatiotemporal registration. Each row corresponds to one temporally evolving 4D tree. Following registration, the trees are better temporally aligned with one another, showing consistent structural development over time. This improved synchronization contrasts with the unregistered examples shown in Fig. 15 in the supplementary material }
    \label{fig:Set1_after_reg}
\end{figure}

\begin{figure}[tb]
    \centering
    \includegraphics[width=\linewidth]{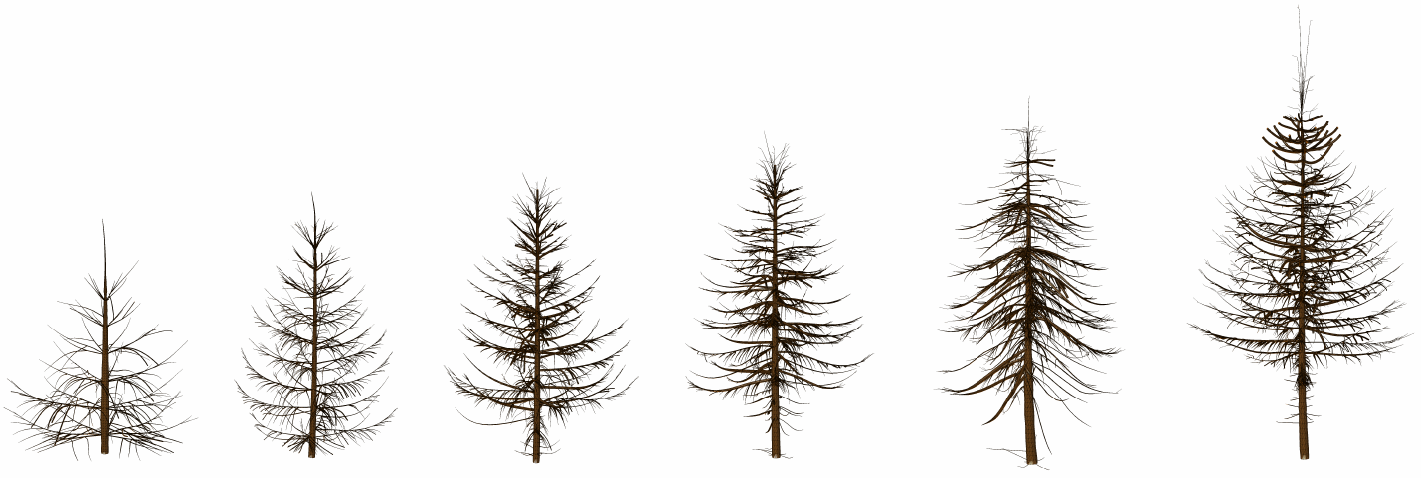}
    \includegraphics[width=\linewidth]{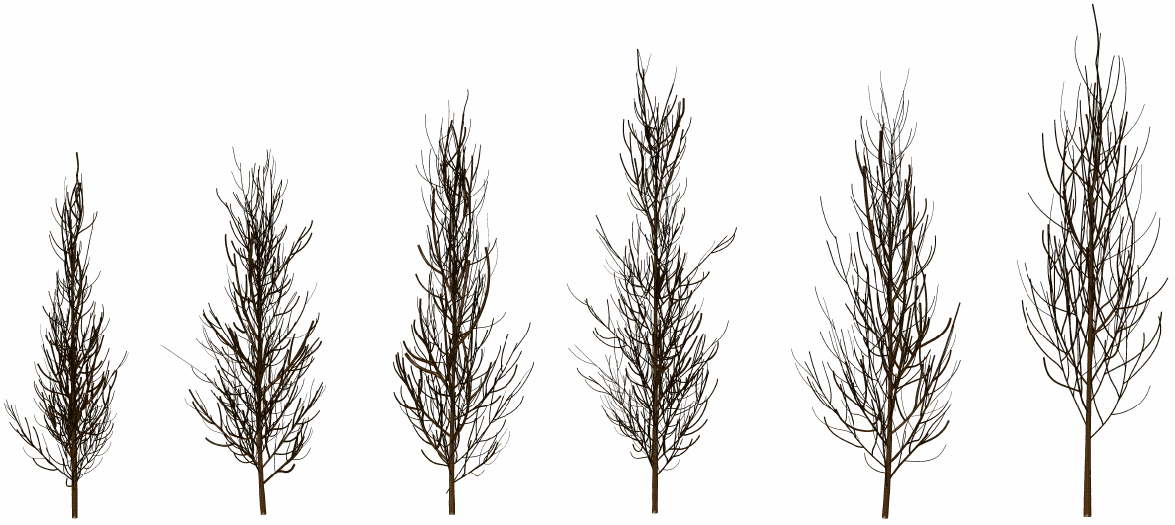}
    \small{\textbf{(a)} Set 2 of 4D tree models.}
    \includegraphics[width=\linewidth]{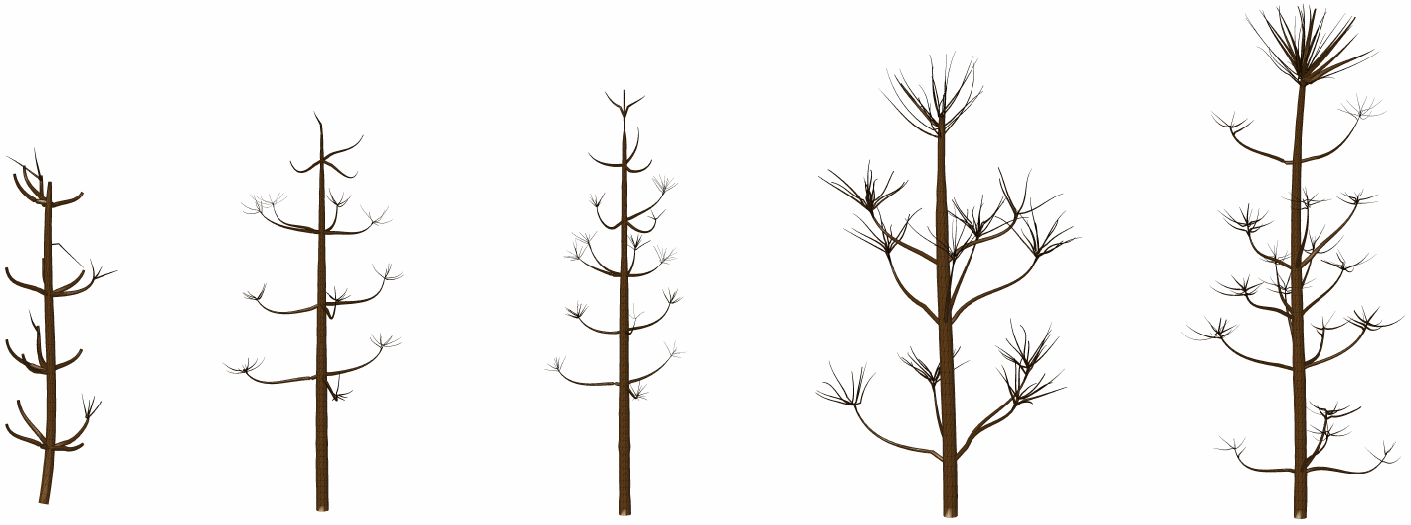}
    \includegraphics[width=\linewidth]{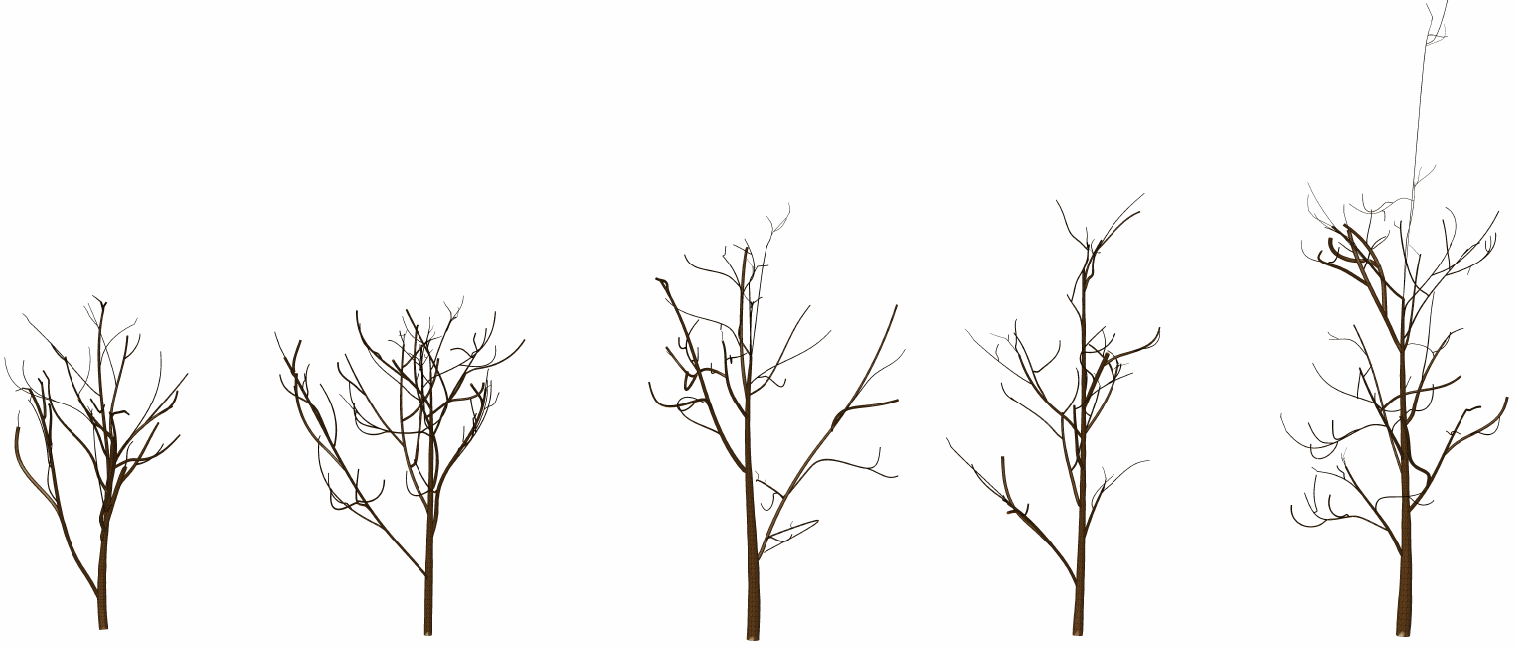}
    \small{\textbf{(b)} Set 3 of 4D tree models.}
    \caption{Visualization of the second \textbf{(a)} and third \textbf{(b)} sets of 4D tree models after applying spatiotemporal registration within each set. Each row depicts the temporal progression of a single 4D tree. The results illustrate improved temporal alignment of branching structures across time steps within individual trees in each set.} 
    \label{fig:set2and3_after_reg}
\end{figure}

\begin{figure}[tb]
    \centering
    \includegraphics[width=\linewidth]{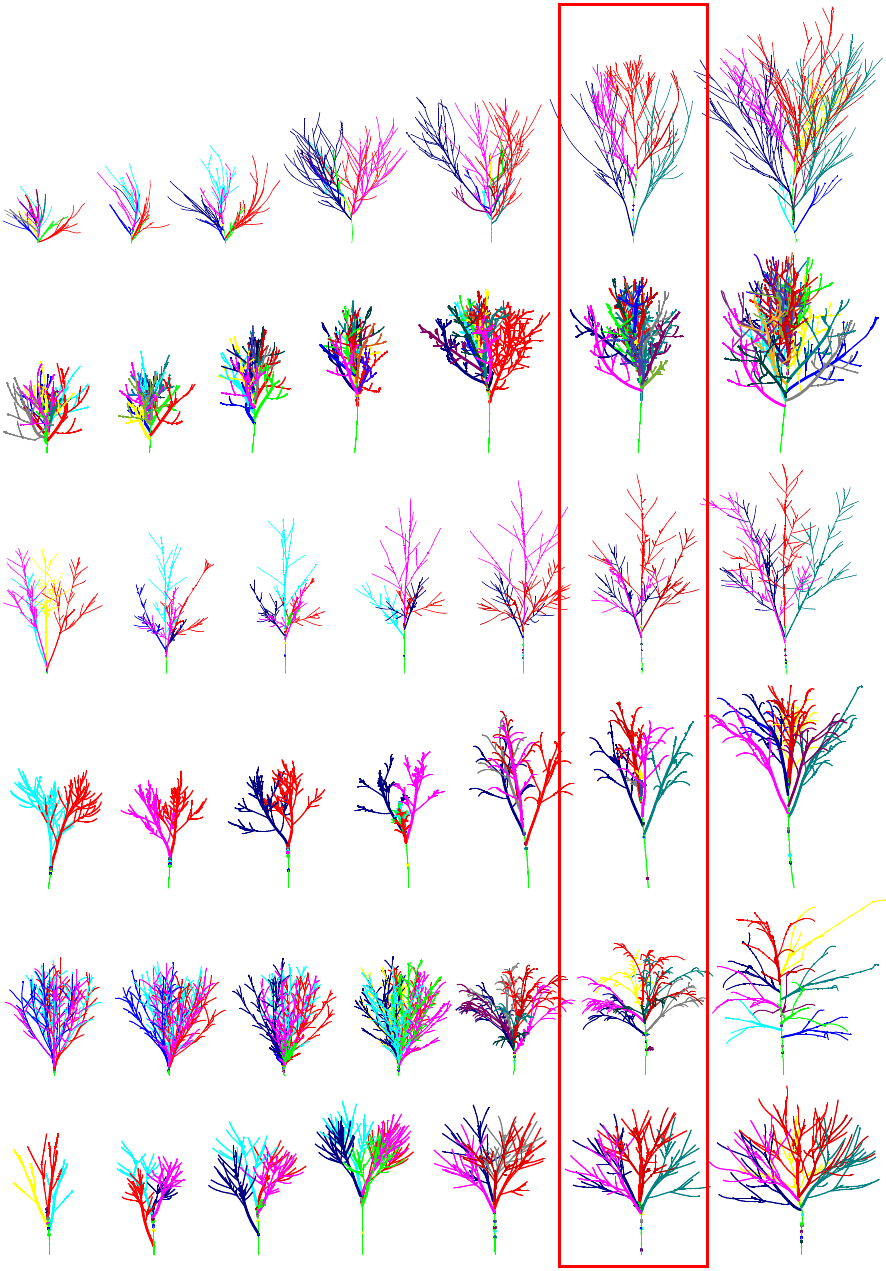}
    \caption{Set 1 of 4D tree models visualized at the skeleton level after applying spatiotemporal registration. Each row corresponds to a single temporally evolving 4D tree. Branch-wise correspondences are color-coded up to the second level of the hierarchy. The red box highlights trees where branch colorings exhibit improved temporal consistency, indicating successful registration. Compared to Fig. 17 in the supplementary material (pre-registration), the color patterns within the red box are more coherent. A zoomed-in view of these trees is provided in Fig. 23 of the supplementary material for better visualization.}
    \label{fig:Set1_after_reg_skel}
\end{figure}

\subsection{Summary statistics and 4D tree-shape synthesis}
\label{sect:summary_statistics}
\begin{figure}[tb]
    \centering
    \includegraphics[width=\linewidth]{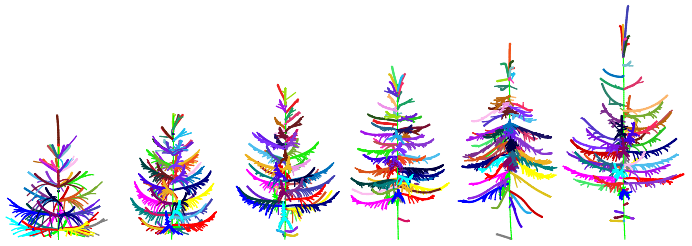}
    \includegraphics[width=\linewidth]{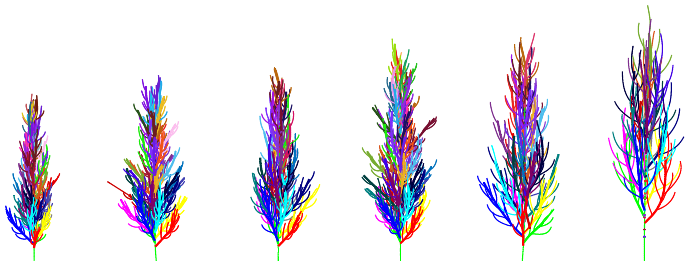}
    \small{\textbf{(a)} Set 2 of 4D tree models at skeleton level, after spatiotemporal registration. }
    \includegraphics[width=\linewidth]{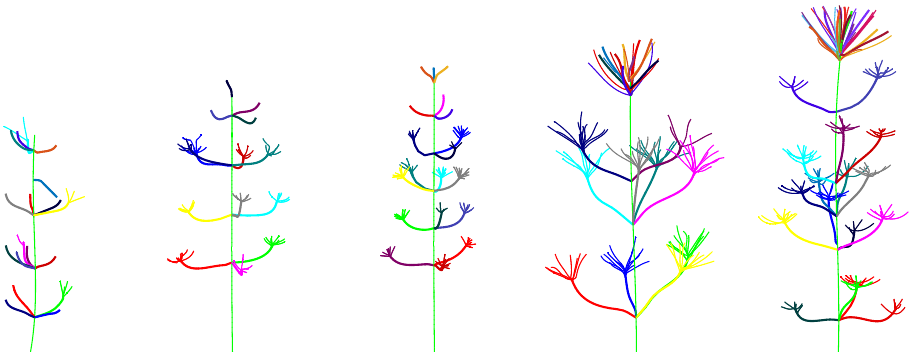}
    \includegraphics[width=\linewidth]{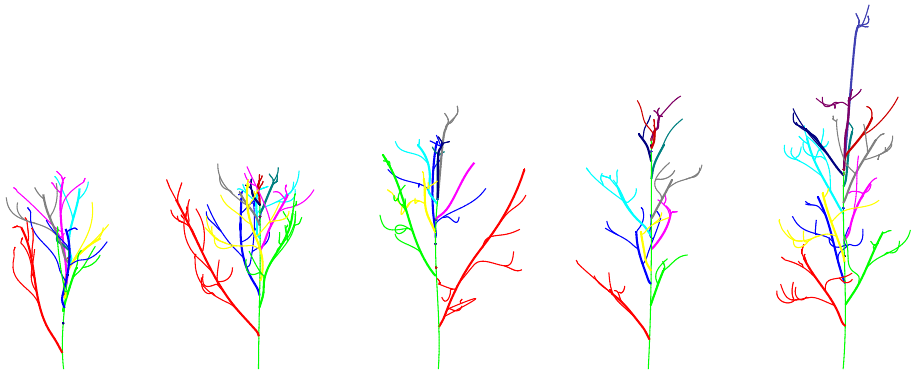}
    \small{\textbf{(b)} Set 3 of 4D tree models,  shown as skeletons, after spatiotemporal registration.}
    \caption{Visualization of \textbf{(a)} Set 2 and \textbf{(b)} Set 3 of 4D tree models shown at the skeleton level following spatiotemporal registration. Each row represents a single 4D tree evolving over time. Branch-wise correspondences are color-coded up to the second level of the hierarchy. The results demonstrate improved internal consistency of branch alignments within each set, compared to their unregistered counterparts shown in Fig. 18 in the supplementary material.
    }
    \label{fig:set2and3_after_reg_skel}
\end{figure}

We take ten unregistered 4D tree models from \href{https://globeplants.com}{Globe plants} and group them into three different sets based on their structural similarity; see  
Figs. 15 and 16 in the supplementary material. We then use our framework to co-register the 4D tree models within each set. Figs.~\ref{fig:Set1_after_reg} and~\ref{fig:set2and3_after_reg} show those sets after their spatiotemporal registration.
We also show the branch-wise correspondences in these sets before (Figs. 17 and 18 in the supplementary material) and after co-registration (Figs.~\ref{fig:Set1_after_reg_skel} and~\ref{fig:set2and3_after_reg_skel}).

\begin{figure}[!tb]
    \centering
    \includegraphics[width=\linewidth]{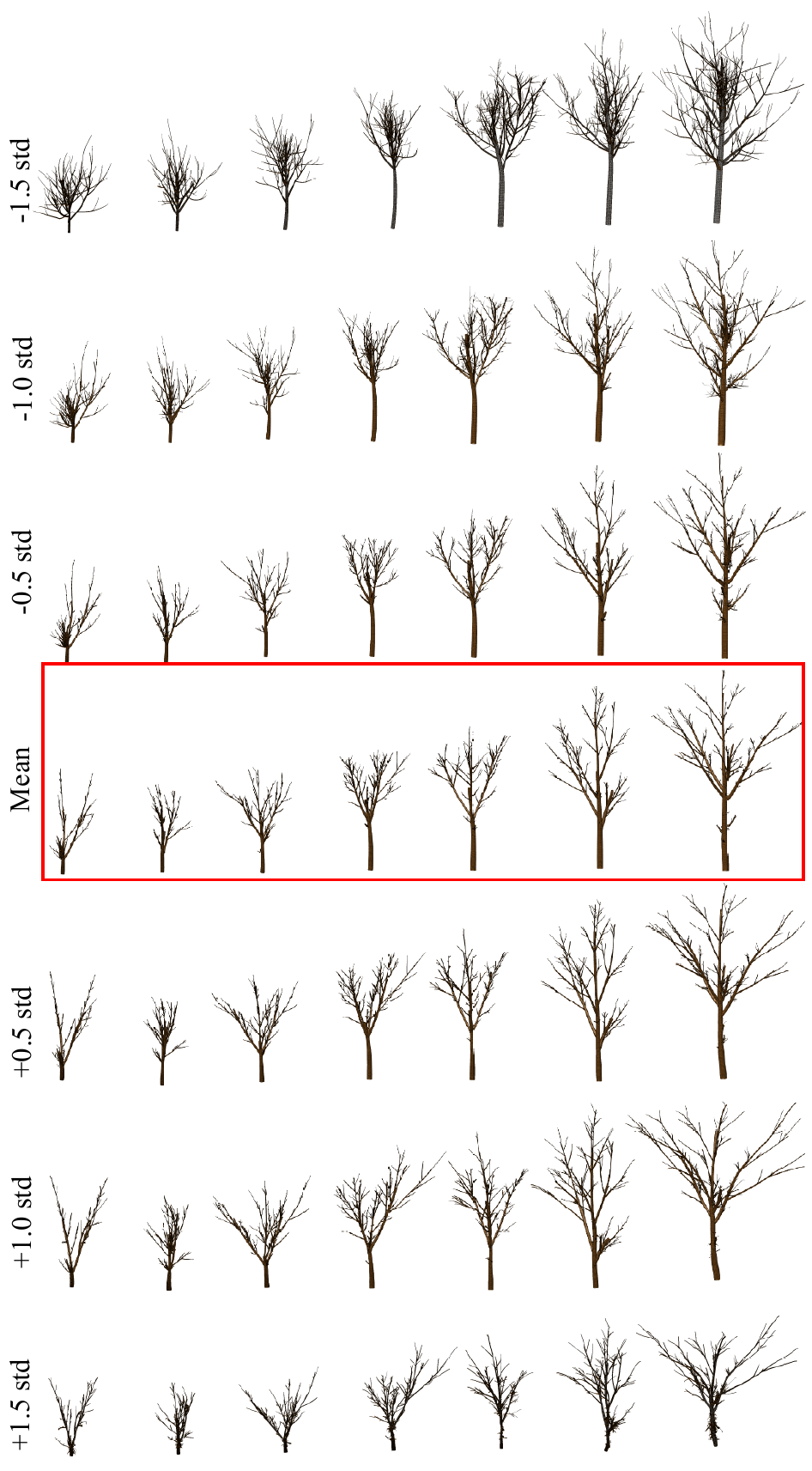}
    \caption{Visualization of the mean 4D tree (highlighted in red) and variations along the first principal component direction, computed from the 4D trees in Set 1 (see Fig.~\ref{fig:Set1_after_reg}). Each row corresponds to a trajectory sampled at increasing standard deviations  from the mean along the first mode of variation. These results illustrate how the primary source of spatiotemporal variability affects overall tree growth and structure.}
    \label{fig:mean_mode_set1}
\end{figure}

\begin{figure}[!tb]
    \centering
    \includegraphics[width=\linewidth]{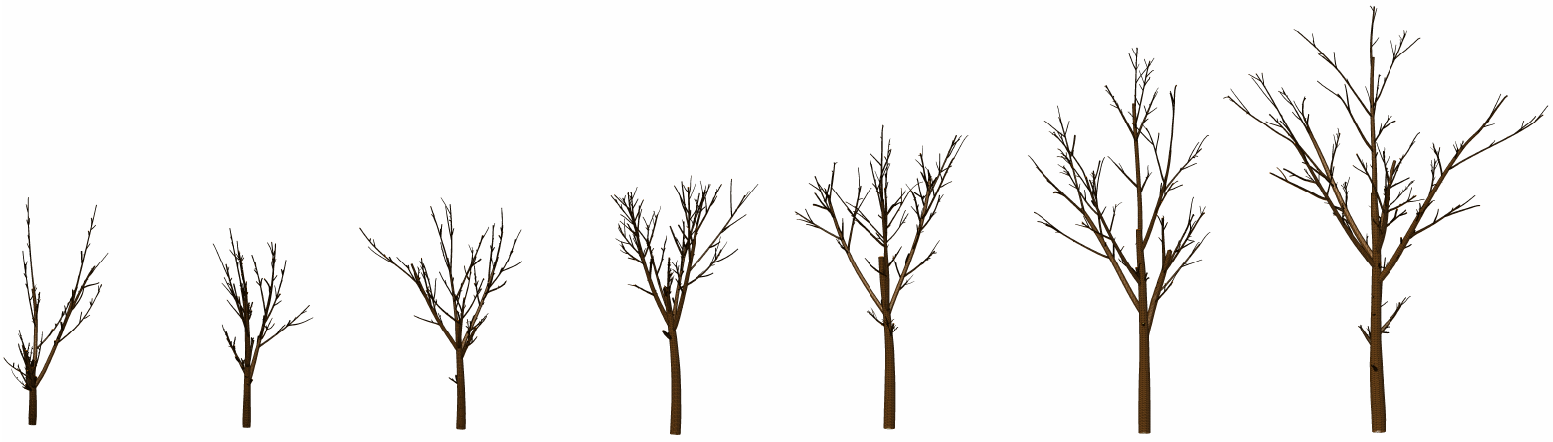}
    \includegraphics[width=\linewidth]{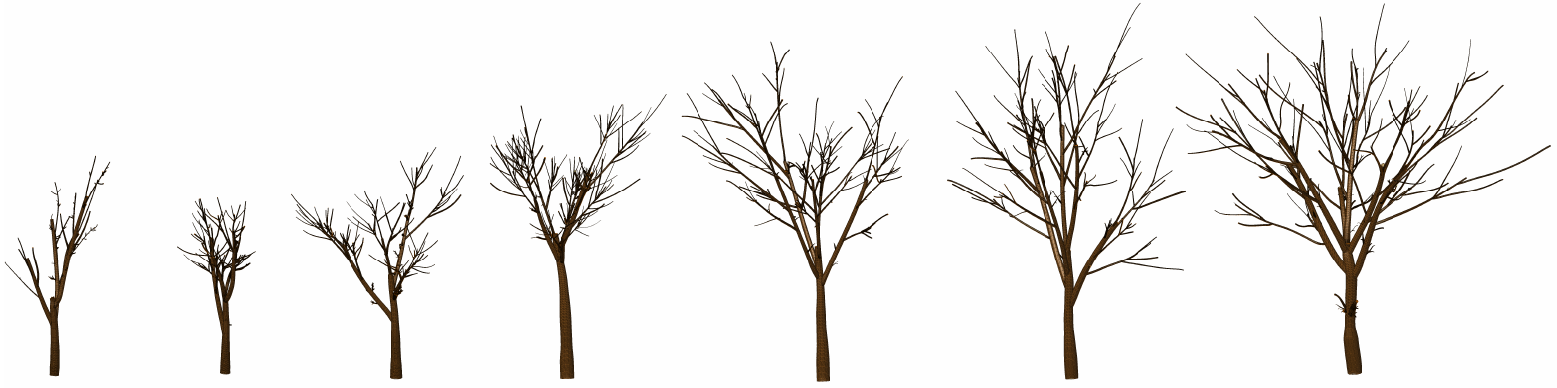}
    \includegraphics[width=\linewidth]{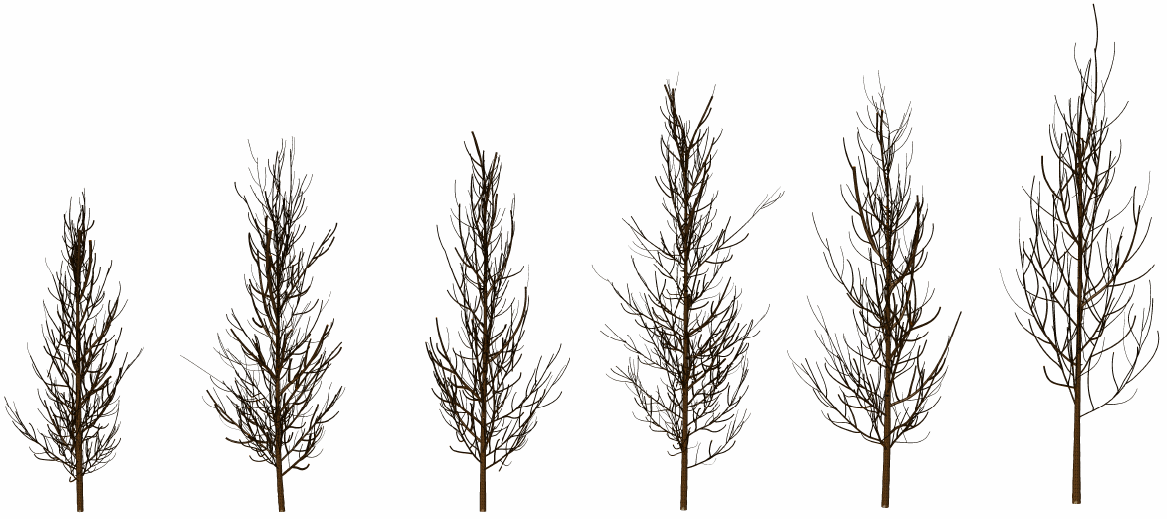}
    \includegraphics[width=\linewidth]{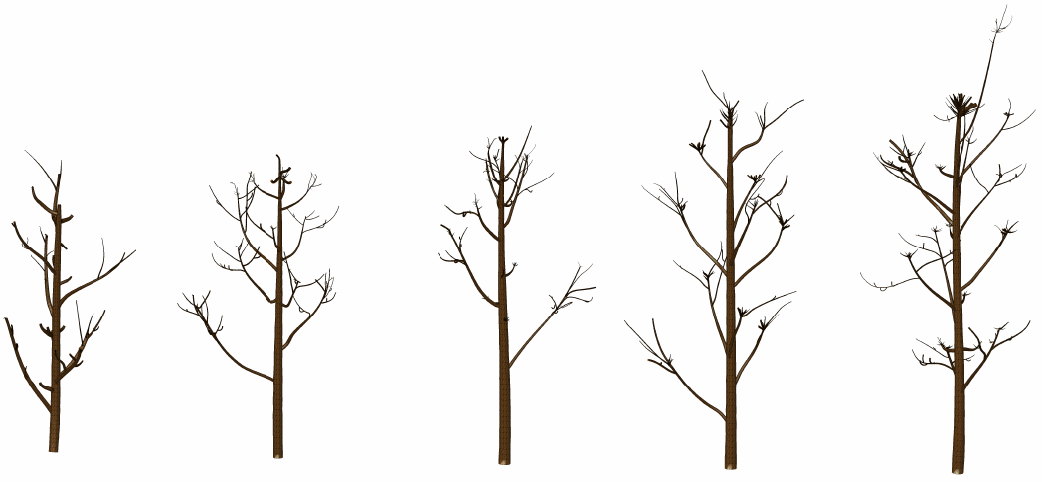}
    \caption{Randomly synthesized 4D tree models, with one 4D tree shown per row. The top two rows are generated from the learned distribution of 4D trees in Set 1 (see Fig.~\ref{fig:Set1_after_reg}). The third row is generated using 4D trees from Set 2 (see Fig.~\ref{fig:set2and3_after_reg}(a)), and the fourth row is based on 4D trees from Set 3 (see Fig.~\ref{fig:set2and3_after_reg}(b)). These results demonstrate the generative capability of the proposed framework in capturing the structural and temporal variability of tree-like 4D shapes.}
    \label{fig:synthesis_results}
\end{figure}

Next,  we compute the mean and modes of variation of these 4D tree models within each subset using the proposed framework. The middle row of Fig.~\ref{fig:mean_mode_set1} shows the mean 4D tree of the registered 4D tree models of Fig.~\ref{fig:Set1_after_reg}. We can see that the mean captures the main characteristics of the 4D trees in the corresponding set.
The figure also shows the first mode of variation, \ie the 4D trees sampled from $-1.5$ to $+1.5$  standard deviation. Figs. 19 and 20 in the supplementary material show additional results on the two sets of Figs.~\ref{fig:set2and3_after_reg}(a) and~\ref{fig:set2and3_after_reg}(b), respectively. 
Fig.~\ref{fig:synthesis_results} shows four randomly generated 4D tree models using our proposed framework. In this experiment, the generative model is learned from the ten 4D tree models. To obtain plausible random 4D trees, we restricted the randomization within $-3$ to $+3$ times the standard deviation along each principal direction of variation. On average, generating one 4D tree takes  $1$ second.

\begin{figure}[!tb]
    \centering
    \begin{tabular}{@{}c@{ }c@{}}
    \includegraphics[width=0.48\linewidth]{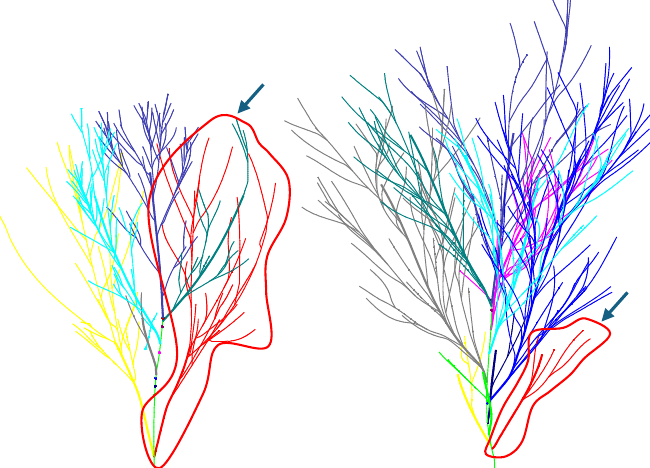} &
    \includegraphics[width=0.48\linewidth]{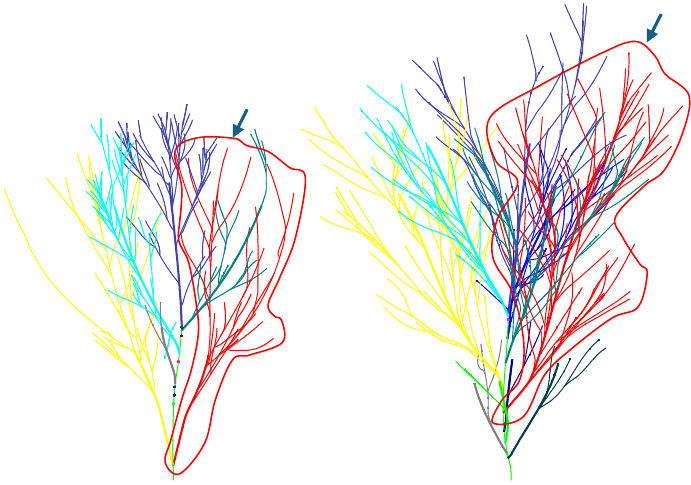}\\
    \small{\textbf{(a)} $\lambda_m=1, \lambda_s=0.1, \lambda_p=1$.} &  
    
    \small{\textbf{(b)} $\lambda_m= \lambda_s=1, \lambda_p=0.5$.}

    \end{tabular}
    \caption{Evaluation of the impact of metric parameter settings on spatial registration outcomes. The color codes indicate branch-wise correspondences up to the second hierarchical level. \textbf{(a)} Results using parameters $\lambda_m=1, \lambda_s=0.1,$ and $\lambda_p=1$, emphasizing minimization of distances between bifurcation points. This results in tighter alignment of branch junctions, as highlighted in red, but potentially at the expense of preserving side-branch structure. \textbf{(b)} Results with parameters $\lambda_m= \lambda_s=1, \lambda_p=0.5$, which balance alignment between bifurcation points and branch geometry, leading to improved side-branch correspondence.}
    \label{fig:effect_of_parameters}
\end{figure}

\subsection{Ablation study}
\label{sect:ablation}
We conduct an ablation study to analyze  \textbf{(1)} how the parameters ($\lambda_m,\lambda_s,\lambda_p$)  affect the spatial registration, \textbf{(2)} the importance of the SRVF representation for temporal registration, \textbf{(3)} how Yeo-Johnson power transformations~\cite{weisberg2001yeo} affect the mapping of tree models into the PCA space, and \textbf{(4)} the importance of the ESRVF representation for spatial correspondence.

\vspace{4pt}
\noindent
\textbf{(1) Effects of the parameters $(\lambda_m,\lambda_s,\lambda_p) \in [0, 1]$.} 
We have varied these parameters to evaluate their effect on the spatial registration. We observe that these parameters control the spatial registration process. $\lambda_m$  controls the importance of the distance between individual branches, whereas $\lambda_s$ emphasizes the similarity between side branches of a source and target branch. Thus, setting $\lambda_s=1$ will likely favor the alignment of those branches that have similar side branches. If we decrease the value of $\lambda_s$, the spatial registration will focus less on matching the side branches as shown in Fig.~\ref{fig:effect_of_parameters}(a), where we set $\lambda_s=0.1$. On the other hand,  for $\lambda_s=1$, the registration process will focus on the side branch structures and align appropriately, as shown in Fig.~\ref{fig:effect_of_parameters}(b). $\lambda_p$ emphasizes the distance between bifurcation points of the source and target branches. As shown in Fig.~\ref{fig:effect_of_parameters}(a), with $\lambda_s=0.1$ and $\lambda_p=1$,  the registration focuses on minimizing the distance between bifurcation points rather than aligning the side branches. In Fig.~\ref{fig:effect_of_parameters}(b), where we set $\lambda_p=0.5$ and $\lambda_s=1$,  the registration focuses on the similarity between side branches more than on minimizing the distance between the bifurcation points. 

\begin{figure}[tb]
    \centering
    \includegraphics[width=\linewidth]{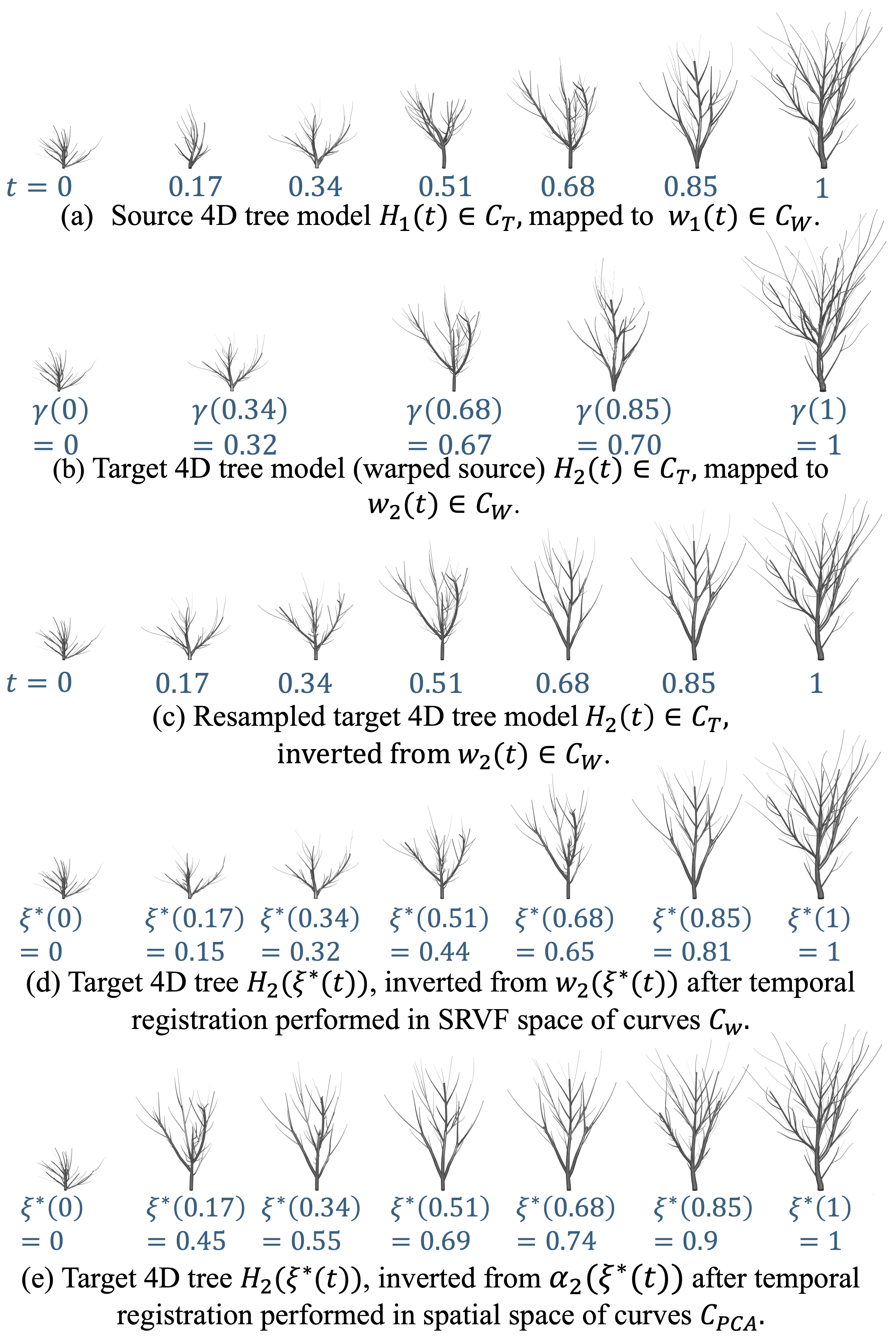}
    \caption{Impact of representation choice on temporal alignment. \textbf{(a)} Source 4D tree. \textbf{(b)} Target 4D tree generated by applying a non-uniform temporal warp $\gamma: [0,1]  \to [0,1]$ to the source. \textbf{(c)} Uniformly resampled target to match the temporal resolution of the source. \textbf{(d)} Result after applying optimal reparameterization $\reparmcurve\optimal$ computed with our method using the SRVF representation, which enables alignment under the elastic metric~\cite{srivastava2010shape}. \textbf{(e)} Result after applying optimal reparameterization $\reparmcurve\optimal$ computed in the original space of curves $\pcaspace$. The SRVF-based approach \textbf{(d)} yields better synchronization of structural growth stages compared to \textbf{(e)}, demonstrating its effectiveness in modeling and correcting temporal distortions.}
    \label{fig:importance_of_srvf}
\end{figure}

\begin{figure} 
    \centering
    \includegraphics[width=\linewidth]{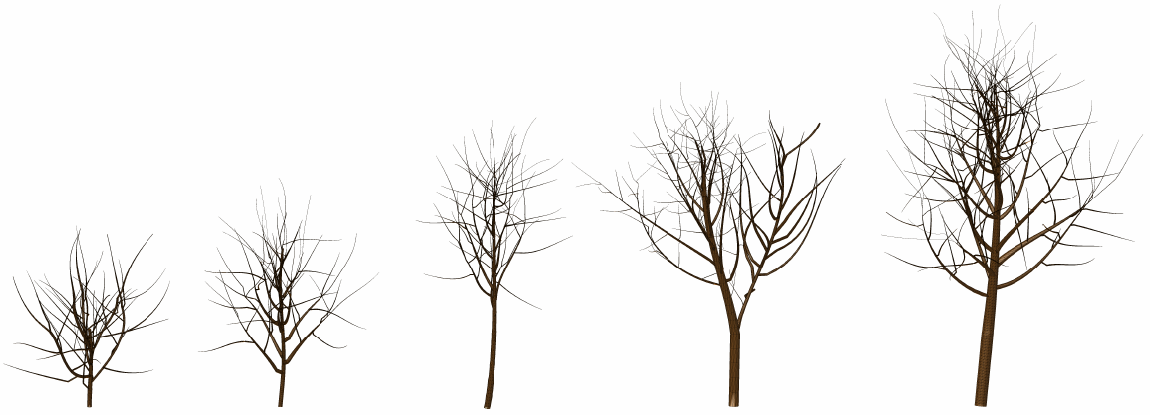}
    \small{\textbf{(a)} Original 4D tree model.}
    \includegraphics[width=\linewidth]{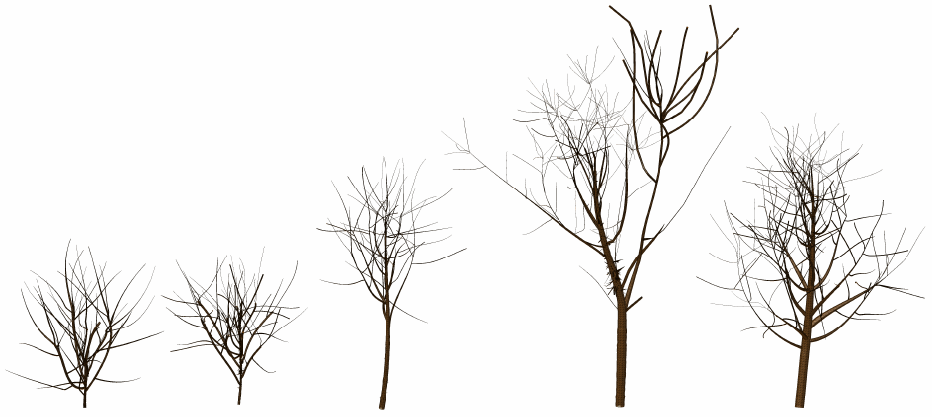}
    \small{\textbf{(b)} Reconstructed 4D tree model from PCA space without applying the  Yeo-Johnson transformation.} 
    \includegraphics[width=\linewidth]{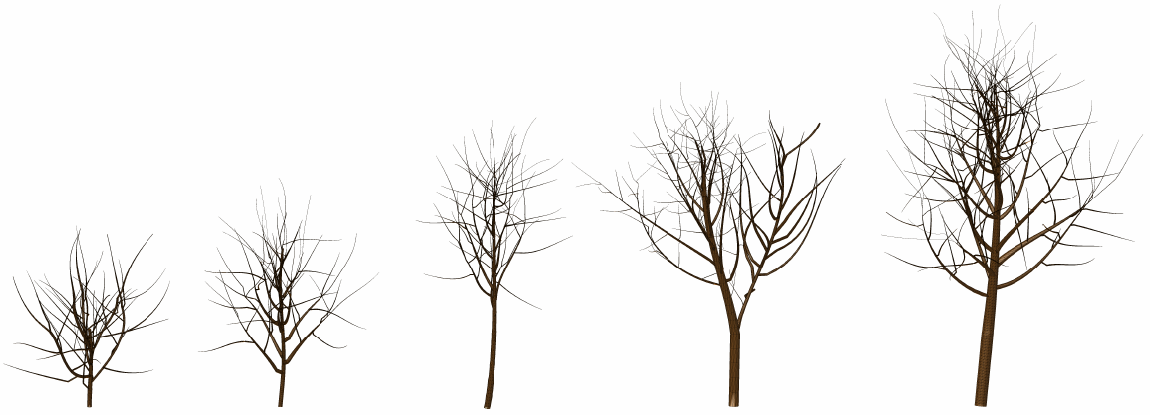}
    \small{\textbf{(c)} Reconstructed 4D tree model from PCA space after applying the  Yeo-Johnson transformation.} 
    \includegraphics[width=\linewidth]{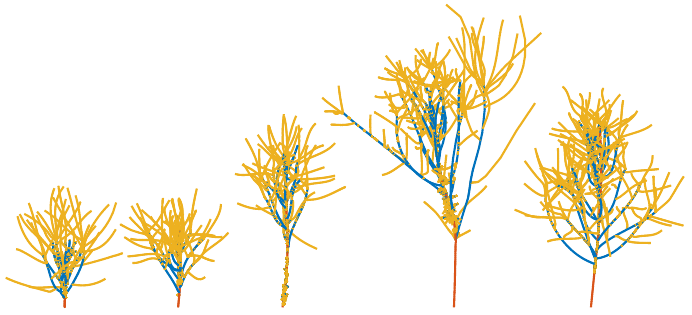}
    \small{\textbf{(d)} The skeleton of the reconstructed 4D tree model of \textbf{(b)}.}
    \includegraphics[width=\linewidth]{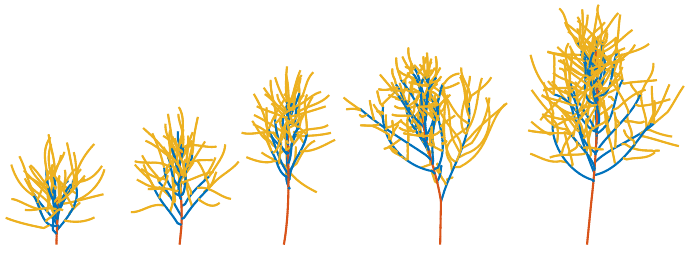}
    \small{\textbf{(e)} The skeleton of the reconstructed 4D tree model of \textbf{(c)}.} 
    \caption{Analysis of the effect of the Yeo–Johnson transformation on PCA-based reconstruction of 4D tree models. \textbf{(a)} Original 4D tree model. \textbf{(b)} Reconstructed model from PCA space without applying the Yeo–Johnson transformation, showing visible structural inconsistencies. \textbf{(c)} Reconstructed model after applying the Yeo–Johnson transformation, yielding a more accurate approximation of the original. \textbf{(d)} and \textbf{(e)} display the corresponding skeletons of \textbf{(b)} and \textbf{(c)}, respectively, further illustrating that the Yeo–Johnson transformation leads to better branch-wise alignment and preservation of structural details.}
    \label{fig:importance_of_Yeo_johnson}
\end{figure}

\begin{figure}[t]
    \centering
    \includegraphics[width=\linewidth]{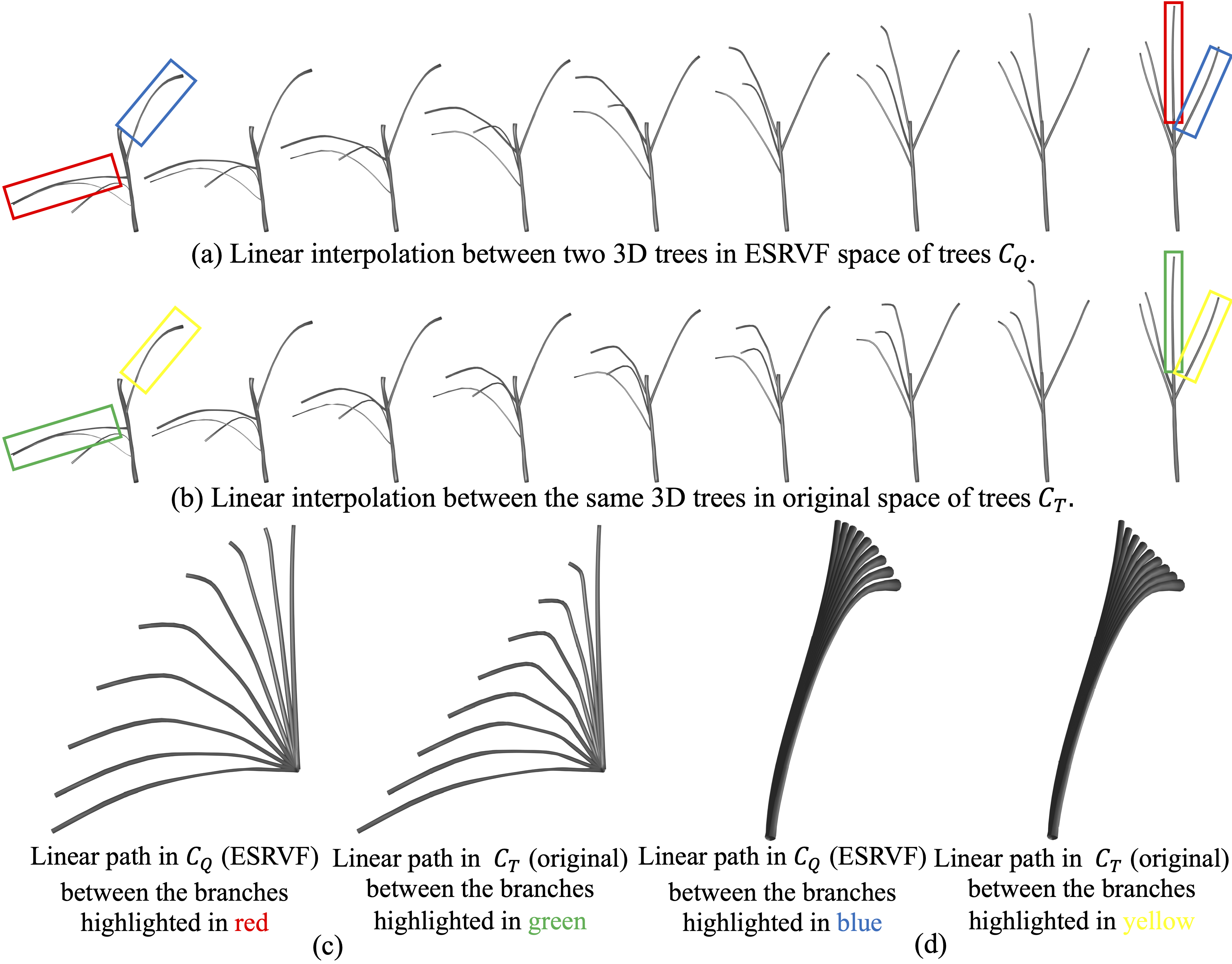}
    \caption{Importance of the proposed ESRVF space and the associated $\ltwo$ metric in the analysis tasks. (a) Linear interpolation between the leftmost and the rightmost 3D trees in the proposed ESRVF-space, $\preshapespaceThreeDTreees$. (b) Linear interpolation between the same 3D trees as (a), but this time in the original tree shape space $\preshapesrvts$. For clarity, we picked two pairs of branches from the two trees and plotted them individually in (c) and (d), respectively. We can clearly see that the linear interpolation in the original space suffers from shrinkage. However, the interpolation becomes natural in our proposed ESRVF space.}
    \label{fig:ESVF_imapct}
\end{figure}

\vspace{4pt}
\noindent
\textbf{(2) Importance of the SRVF representation in temporal registration.} 
We take a ground-truth 4D source tree model and randomly warp it, reducing its temporal resolution, to generate a target 4D tree model. Hence, the target becomes not synchronized with the source. We then register the target 4D tree model onto the source 4D tree model, both in the space of curves $\pcaspace$ and in the SRVF space $\srvfpcaspace$.  Fig.~\ref{fig:importance_of_srvf} shows the source (first row), the warped target (second row), the target after temporal registration performed in the SRVF space (third row), and the target after temporal registration performed in the original space of curves $\pcaspace$ (fourth row). We can observe that, in the original space, the temporal registration failed to synchronize the target with the source. We also observe a significant length distortion in the 4D target model, and thus it cannot properly represent the growth of the target tree. This is expected since the $\ltwo$ metric does not capture nonlinear elastic deformations.  

\noindent
\textbf{(3) Effect of Yeo-Johnson power transformations.} 
Since PCA works well on data that follows a Gaussian distribution, we first apply to the 3D tree models the Yeo-Johnson power transformations~\cite{weisberg2001yeo} to map the original distribution to a  Gaussian distribution. As we can see in Fig.~\ref{fig:importance_of_Yeo_johnson}, when we map the 3D tree models into PCA space without applying the Yeo-Johnson transformation, the inverse PCA adds noise to the original data (see Fig.~\ref{fig:importance_of_Yeo_johnson}(b) and (d)). However, when we perform the Yeo-Johnson transformation first and then apply  PCA, the reconstruction of the original data becomes highly accurate (see Fig.~\ref{fig:importance_of_Yeo_johnson}(c) and (e)). 

\vspace{4pt}
\noindent
\textbf{(4) The importance of the ESRVF representation.} 
Fig.~\ref{fig:ESVF_imapct} illustrates the shrinkage effect when using the $\ltwo$ metric in the original tree shape space $\preshapespaceThreeDTreees$. In this example, we select two sample 3D trees from two different 4D sequences. We show that, even after registration,  linear interpolation in the original space of trees, even when the trees have the same number of branches,  exhibits unnatural shrinkage. In contrast, we can observe that branch bending is more accurately captured when using the $\ltwo$ metric in our proposed ESRVF space $\preshapesrvts$,  leading to a natural deformation without shrinkage.

\section{Conclusion}
\label{sec:conclusion}
This paper introduced a comprehensive mathematical framework for the statistical analysis of 4D tree-shaped structures, such as plants and botanical trees, which exhibit complex geometries and intricate branching patterns. We proposed a novel representation that models 4D trees as trajectories in a Riemannian shape space, where the adoption of the $\ltwo$ metric becomes equivalent to employing a full elastic metric~\cite{jermyn2012elastic}. This reformulation substantially simplifies the analysis by reducing it to the study of curves in a Euclidean space. Using this framework, we demonstrated its effectiveness in several key tasks: spatiotemporal registration of 4D tree-like objects, computation of geodesics between 4D trees, analysis of spatiotemporal variability across shape collections, and the estimation of probability distributions over these structures. These distributions further enable the synthesis of realistic, randomly generated 4D tree-shaped objects. While our experimental validation focused primarily on botanical trees and plants, the proposed framework is broadly applicable to other 4D tree-like data, including neuronal structures in the brain, vascular and airway systems in the human body, and natural branching phenomena such as river networks. Importantly, our approach is grounded in physical principles and does not rely on deep learning. As a promising direction for future work, we aim to investigate how the proposed physics-based metric can be incorporated into modern deep learning pipelines, enabling learning-based generative models that exploit the continuous latent structure defined by our framework.

\bibliographystyle{IEEEtran}
\bibliography{main}

\end{document}